\documentclass[reprint,amsmath,amssymb, aps,pre,floatfix,superscriptaddress]{revtex4-1}

\usepackage[hidelinks]{hyperref}
\usepackage{graphicx} 
\usepackage{siunitx}
\usepackage[version=4]{mhchem}
\usepackage{comment}
\usepackage{xcolor}
\usepackage{tikz}
\usepackage{scalerel}
\usepackage{xifthen}
\usepackage{chngcntr}
\usetikzlibrary{svg.path}

\newcommand{\figref}[2][{}]{Fig.\ \ref{#2}\ifthenelse{\isempty{#1}}{}{\,(#1)}}
\newcommand{\plb}[1]{(\MakeLowercase{#1})} 
\renewcommand{\vec}{\mathbf}

\definecolor{orcidlogocol}{HTML}{A6CE39}
\tikzset{
  orcidlogo/.pic={
    \fill[orcidlogocol] svg{M256,128c0,70.7-57.3,128-128,128C57.3,256,0,198.7,0,128C0,57.3,57.3,0,128,0C198.7,0,256,57.3,256,128z};
    \fill[white] svg{M86.3,186.2H70.9V79.1h15.4v48.4V186.2z}
                 svg{M108.9,79.1h41.6c39.6,0,57,28.3,57,53.6c0,27.5-21.5,53.6-56.8,53.6h-41.8V79.1z M124.3,172.4h24.5c34.9,0,42.9-26.5,42.9-39.7c0-21.5-13.7-39.7-43.7-39.7h-23.7V172.4z}
                 svg{M88.7,56.8c0,5.5-4.5,10.1-10.1,10.1c-5.6,0-10.1-4.6-10.1-10.1c0-5.6,4.5-10.1,10.1-10.1C84.2,46.7,88.7,51.3,88.7,56.8z};
  }
}

\newcommand\orcid[1]{\href{https://orcid.org/#1}{\mbox{\scalerel*{
\begin{tikzpicture}[yscale=-1,transform shape]
\pic{orcidlogo};
\end{tikzpicture}
}{|}}}}

\begin{document}

\title{Emergence of bimodal motility in active droplets}
\author{Babak Vajdi Hokmabad~\orcid{0000-0001-5075-6357}}
\affiliation{Max Planck Institute for Dynamics and Self-Organization, Am Fa\ss{}berg 17, 37077 G\"ottingen}
\affiliation{Institute for the Dynamics of Complex Systems, Georg August Universit\"at G\"ottingen}
\author{Ranabir Dey~\orcid{0000-0002-0514-7357}}
\affiliation{Max Planck Institute for Dynamics and Self-Organization, Am Fa\ss{}berg 17, 37077 G\"ottingen}
\affiliation{Institute for the Dynamics of Complex Systems, Georg August Universit\"at G\"ottingen}
\affiliation{Department of Mechanical and Aerospace Engineering, Indian Institute of Technology Hyderabad, Kandi, Sangareddy, Telengana- 502285, India}
\author{Maziyar Jalaal~\orcid{0000-0002-5654-8505}}
\affiliation{Physics of Fluids Group, Max Planck Center for Complex Fluid Dynamics, MESA+ Institute and J. M. Burgers Center for Fluid Dynamics, University of Twente, PO Box 217,7500 AE Enschede, The Netherlands}
\affiliation{Department of Applied Mathematics and Theoretical Physics,
University of Cambridge, Cambridge CB3 0WA, United Kingdom}
\affiliation{Van der Waals-Zeeman Institute, Institute of Physics, University of
Amsterdam, Amsterdam, The Netherlands}
\author{Devaditya Mohanty~\orcid{0000-0001-6797-5206}}
\affiliation{Indian Institute of Technology Guwahati, Assam 781039}
\author{Madina Almukambetova~\orcid{0000-0003-3223-5392}}
\affiliation{Ulsan National Institute of Science and Technology (UNIST), Ulsan 44919, Republic of Korea}
 \author{Kyle A. Baldwin~\orcid{0000-0001-9168-6412}}
\affiliation{Max Planck Institute for Dynamics and Self-Organization, Am Fa\ss{}berg 17, 37077 G\"ottingen}
 \affiliation{Institute for the Dynamics of Complex Systems, Georg August Universit\"at G\"ottingen}
  \affiliation{SOFT group, School of Science and Technology, Nottingham Trent University, Nottingham, NG11 8NS, United Kingdom}
 \author{Detlef Lohse~\orcid{0000-0003-4138-2255}}
\affiliation{Physics of Fluids Group, Max Planck Center for Complex Fluid Dynamics, MESA+ Institute and J. M. Burgers Center for Fluid Dynamics, University of Twente, PO Box 217,7500 AE Enschede, The Netherlands}
 \author{Corinna C. Maass~\orcid{0000-0001-6287-4107}}
 \email{corinna.maass@ds.mpg.de}%
\affiliation{Max Planck Institute for Dynamics and Self-Organization, Am Fa\ss{}berg 17, 37077 G\"ottingen}
\affiliation{Institute for the Dynamics of Complex Systems, Georg August Universit\"at G\"ottingen}
\affiliation{Physics of Fluids Group, Max Planck Center for Complex Fluid Dynamics, MESA+ Institute and J. M. Burgers Center for Fluid Dynamics, University of Twente, PO Box 217,7500 AE Enschede, The Netherlands}

\begin{abstract}

Artificial model swimmers offer a platform to explore the physical principles enabling biological complexity, for example, multi-gait motility: a strategy employed by many bio-microswimmers to explore and react to changes in their environment. 
Here, we report bimodal motility in autophoretic droplet swimmers, driven by characteristic interfacial flow patterns for each propulsive mode. 
We demonstrate a dynamical transition from quasi-ballistic to bimodal chaotic propulsion by controlling the viscosity of the environment. 
To elucidate the physical mechanism of this transition, we simultaneously visualize hydrodynamic and chemical fields and interpret these observations by quantitative comparison to established advection-diffusion models.
We show that, with increasing viscosity, higher hydrodynamic modes become excitable and the droplet recurrently switches between two dominant modes due to interactions with the self-generated chemical gradients.
This type of self-interaction promotes self-avoiding walks mimicking examples of efficient spatial exploration strategies observed in nature.
\end{abstract}

\maketitle

In response to physical constraints in nature, microorganisms have adapted and developed various locomotion strategies. Depending on cues from the environment, these strategies range from the more commonplace helical swimming~\cite{bearon2013_helical,rossi2017_kinematics}, run-and-tumble, and switch-and-flick motility~\cite{stocker2011_reverse}, to more sophisticated transient behaviours, e.g.\ peritrichous bacteria switching poles in response to a steric stress~\cite{cisneros2006_reversal}, octoflagellate microalgae exhibiting run-stop-shock motility with enhanced mechanosensitivity~\cite{wan2018_time}, and starfish larvae maximising fluid mixing, and thereby nutrition uptake, through rapid changes of ciliary beating patterns~\cite{gilpin2017_vortex}.
Such intricate gait-switching dynamics~\cite{tsang2018_polygonal,son2013_bacteria} enable organisms to navigate in external flows~\cite{mathijssen2019_oscillatory,figueroa-morales2020_e}, to follow gradients~\cite{wadhams2004_making} or to efficiently explore their environment~\cite{perezipina2019_bacteria,guadayol2017_cell}.
Recent efforts in the development of synthetic swimmers have led to synthesis of systems that are capable of mimicking some of the aforementioned features of their natural counterparts such as rheotaxis ~\cite{palacci2015_artificial,katuri2018_cross-stream}, chemotaxis~\cite{jin2017_chemotaxis,liebchen2018_synthetic}, and gravitaxis~\cite{tenhagen2014_gravitaxis}.
However, dynamic multi-modal motility in the absence of external actuation has not been explored before in artificial swimmers, and the mechanisms underlying unsteady behaviour in self-actuating systems are not well understood, particularly with respect to distinguishing states with increased random fluctuation from ones featuring true multimodal behavior\cite{suga2018_self-propelled,izzet2020_tunable}.

Paradigms for biomimetic artificial swimmers include autophoretic microswimmers, powered by chemical activity at their interface, which are able to generate long-living chemical gradients in the environment~\cite{liebchen2018_synthetic}.
In this regard, droplet microswimmers driven by micellar solubilization~\cite{maass2016_swimming}, provide a sophisticated experimental realisation. 
Unlike most synthetic swimmers which are inherently asymmetric, active droplets are isotropic. Interfacial activity spontaneously breaks the symmetry, allowing for emergence of different flow patterns depending on the environmental parameters. Here we use such active droplets as model systems to demonstrate the physical principles guiding the emergence of multi-modal motility in response to changes in environmental conditions.

We show that active droplets adapt to an increase in the viscosity of the swimming medium by exhibiting increasingly chaotic motion -- a counter-intuitive response given that increasing viscous stress generally tends to stabilise non-inertial dynamics. Using time-resolved \textit{in situ} visualisation of the chemical and the hydrodynamic fields around the droplet interface, we found that the emergence of the chaotic dynamics correlates with the onset of higher hydrodynamic modes at increasing P\'eclet number $Pe$. 
Once these higher modes prevail, the droplet exhibits an unsteady bimodal exploration of space triggered by its interaction with a self-generated, slowly-decaying chemical gradient. The conditions for the onset of this dynamical transition are quantitatively predicted by an advection-diffusion model for the transport of the chemical species, which takes into account the nonlinear coupling between the hydrodynamic and chemical fields.
The visualisation technique and the findings presented here lay the groundwork for future investigations of emergent dynamics in active phoretic matter.

\section*{Droplets propelled by micellar solubilisation}
Our experiments use a specific subclass of active droplets: oil droplets that are slowly dissolving in supramicellar aqueous solutions  of ionic surfactants. The droplets spontaneously develop self-sustaining gradients in interfacial surfactant coverage, resulting in Marangoni stresses which lead to self-propulsion~\cite{herminghaus2014_interfacial}.
This interfacial instability may be understood as follows (\figref[a,b]{fig:fillingcartoon}):
During the solubilisation of the droplet, oil molecules migrate into surfactant micelles in a boundary layer around the droplet interface, causing the micelles to swell and take up additional surfactant monomers from the aqueous phase, therefore reducing the local density of monomers $c$ below its equilibrium, the critical micelle concentration (CMC).
Unless there are empty micelles present to restore the CMC by disintegration, this local mismatch will reduce the interfacial surfactant coverage, such that the interfacial tension increases with the local ratio of filled to empty micelles.
Following an advective perturbation in the vicinity of the droplet, the initially radially isotropic cloud of filled micelles is distorted; the resulting fore-aft asymmetry generates a surface tension gradient towards the trailing oil-filled micelles which drives the droplet forward. Due to this self-sustaining gradient, the droplet propels continuously,  while leaving behind a trail of swollen micelles (\figref[d]{fig:fillingcartoon}). 

As proposed by hydrodynamic theory models~\cite{michelin2013_spontaneous,izri2014_self-propulsion,morozov2019_nonlinear,morozov2019_self-propulsion,morozov2020_adsorption}, such spontaneous self-propulsion stemming from an advection-diffusion driven interfacial instability arises only if the  P\'eclet number, $Pe$, which characterises the ratio of advective to diffusive transport, exceeds a critical threshold. 
In a simplified description, the surfactant dynamics are approximated by treating the droplet interface as a sink for surfactant monomers~\cite{michelin2013_spontaneous,izri2014_self-propulsion,morozov2019_nonlinear}. In this framework, on which we will base our subsequent mode stability analysis, $Pe$ can be shown to be a monotonically increasing function of the swimming medium ({\it outer}) viscosity $\mu^o$, here non-dimensionalised as $\mu=\mu^o/\mu^i$ using the constant {\it inner} viscosity $\mu^i$ (see Appendix \ref{sisec:peclet} for a step-by-step derivation of eqn.~\eqref{eqn:pe}):
\begin{equation} \label{eqn:pe}
Pe=\frac{V_t R_d}{D} \approx \frac{18 \pi^2}{k_BT}  q_s r_s^2 \zeta R_d^2 \mu^i \left[\mu \left(\frac{2\mu+3 \zeta/R_d}{2 \mu+3}\right)\right],
\end{equation}
where $V_t$ is the theoretical terminal droplet velocity in a
surfactant gradient~\cite{anderson1989_colloid,morozov2019_nonlinear}, $R_d=30\,\mu$m the droplet radius, $D=\frac{k_BT}{6 \pi r_s \mu^o}$ the diffusion coefficient for the surfactant monomer (length scale $r_s \sim 10^{-10}$\,m), $q_s$ the isotropic interfacial surfactant consumption rate per area, and $\zeta \sim 10$\,nm the characteristic length scale over which the surfactants interact with the droplet~\cite{anderson1989_colloid,izri2014_self-propulsion}. 
Increasing $\mu^o$ corresponds to an increase in $Pe$, besides the increase in viscous stresses. Henceforth, we represent an increase in $\mu^o$ by this corresponding increase in $Pe$, as tabulated by the colour map in \figref{fig:statistics}. We note that in view of the necessary simplifications in the derivation of \eqref{eqn:pe}, all experimental $Pe$ values should be regarded as approximate.

In experiments, we controlled $\mu^o$ via water/glycerol mixtures as the swimming medium (viscosity values in Appendix~\ref{appsec:materials}, \figref{sifig:Viscosity}), varying the glycerol content from 0 to 70 vol.\%, and adding the surfactant TTAB at 5 wt\% to generate activity. Monodisperse CB15 oil droplets of radius $R_d=\SI{30}{\micro\metre}$ were studied in quasi-2D reservoirs \SI{60}{\micro\metre} in depth for 5-6 minutes, a time over which the droplet radius should not shrink by more than $\sim$7\%. Therefore, we do not consider any droplet size effects on $Pe$ (see also the experimental materials and methods section in Appendix~\ref{appsec:materials}).

\begin{figure}
\includegraphics[width=\columnwidth]{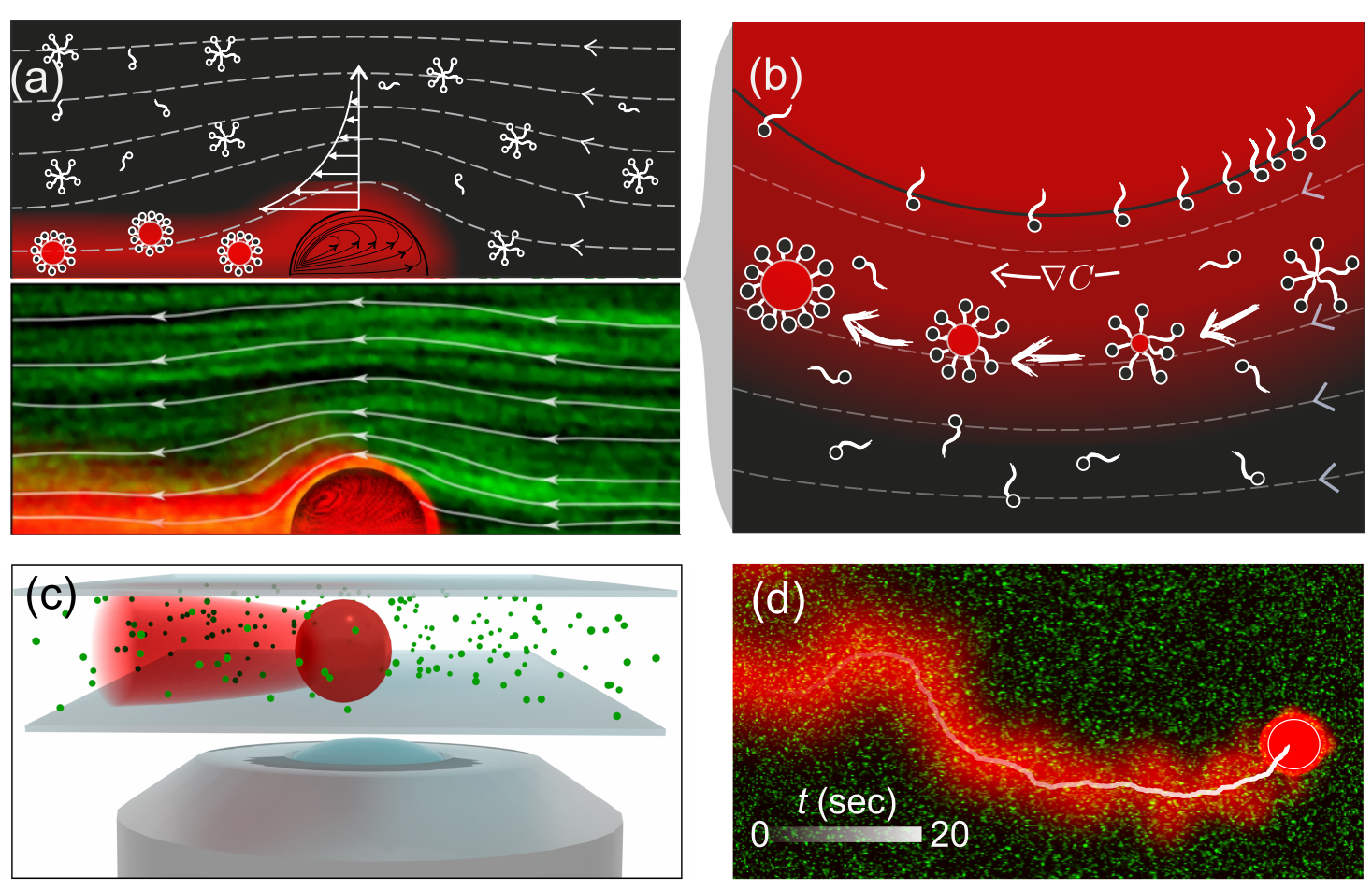}
\caption{\label{fig:fillingcartoon} Droplet propulsion mechanism and visualisation technique. \plb{a} Top: Schematic illustration of the micellar solubilization of oil at the droplet interface leading to self-propulsion.
 Bottom: Streaks of tracers following the flow inside and outside of the droplet during 2 seconds, with streamlines of the external flow from PIV analysis (droplet reference frame). Data from double channel fluorescence microscopy, with illumination at 561\,nm (Nile Red doped oil, red emission) and 488\,nm (tracer colloids, green emission).
 \plb{b} Sketch of the filling and growth of micelles travelling in a boundary layer along the interface, causing a propulsive Marangoni flow. \plb{c} Microscopy set-up schematic with the droplet (radius 30\,$\mu$m) swimming in a Hele–Shaw cell (height 60\,$\mu$m). \plb{d} Sample micrograph, with the droplet's centroid trajectory traced in white.}
\end{figure}

\section*{Simultaneous visualisation of chemical and hydrodynamic fields}
To visualise the chemical and hydrodynamic fields involved in the droplet activity, we directly imaged the chemical field of swollen micelles by adding the hydrophobic dye Nile Red to the oil phase (\figref[c,d]{fig:fillingcartoon}, see also Appendix \ref{sisec:DCFM} and Video S1).
The dye co-migrates with the oil molecules into the filled micelles, which fluoresce when illuminated.
We seeded the surrounding medium, a supramicellar aqueous surfactant solution, with green fluorescent tracer colloids and measured the flow field using particle image velocimetry (PIV).
The emission spectra of dye and colloids are sufficiently non-overlapping to be separately detected in dual channel fluorescence microscopy. 
Consequently, both fields can be simultaneously observed and analysed; 
we provide an example micrograph with an overlay of the extracted droplet trajectory in \figref[d]{fig:fillingcartoon}. Due to the large size ($\sim$ 5\,nm) of the filled micelles, the time scale of their diffusive relaxation exceeds that of the droplet motion; thus, there is a persistent fluorescent trail in the wake of the droplet. 
\begin{figure*}
\includegraphics[width=0.75\textwidth]{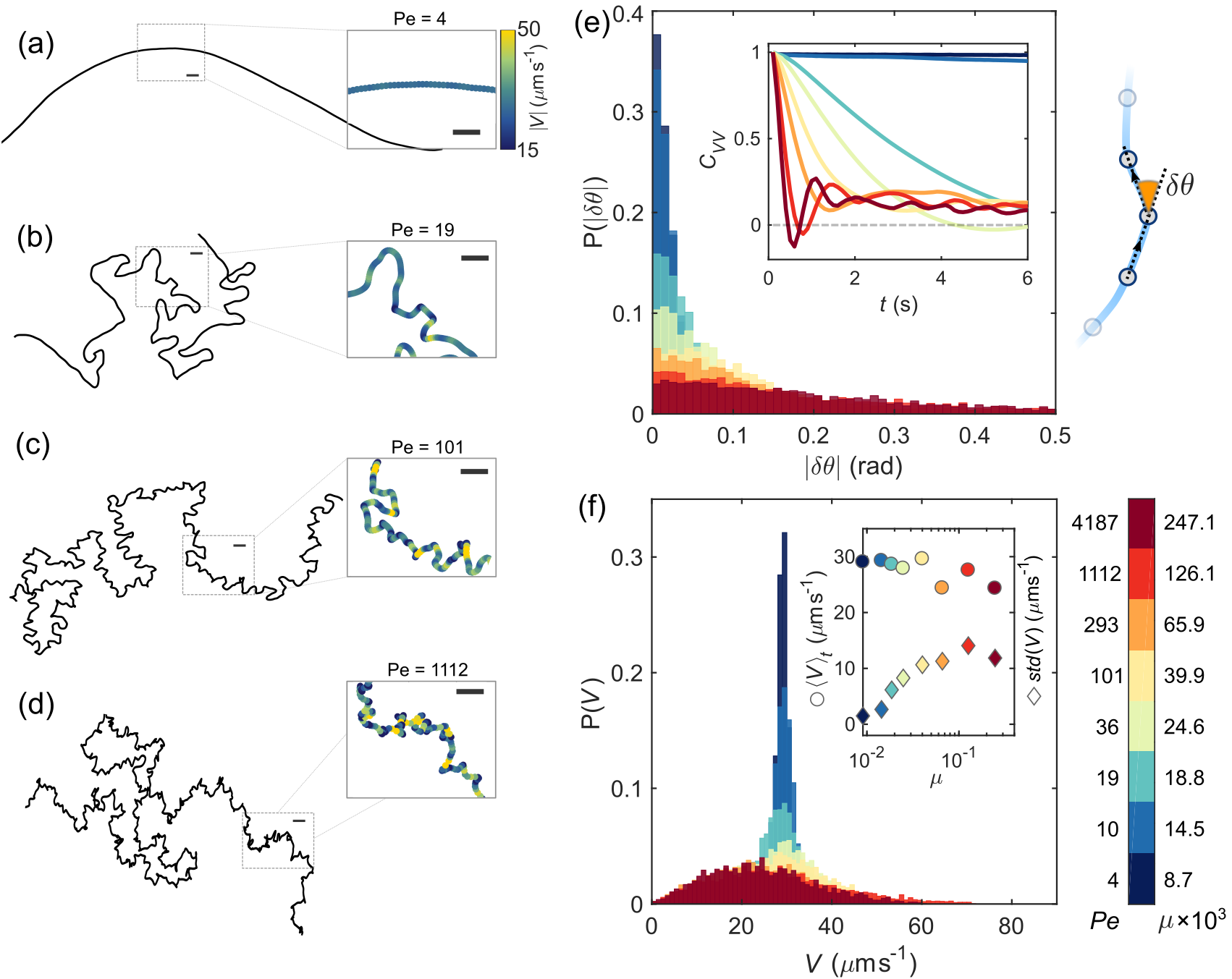}
\caption{\label{fig:statistics} 
Destabilized droplet motion with increasing P\'eclet number $Pe$. \plb{a-d} Example trajectories of droplets for $Pe\in\{4,19,101,1112\}$, with zoomed-in insets colour-coded by propulsion speed $V$. All scale bars are 50\,$\mu$m. \plb{e} Distribution of the velocity reorientation angle, $|\delta\theta|$ for increasing $Pe$, measured during a time step $\delta t=0.1$\,s set by the video recording rate of $10$\,Hz. Profiles of the velocity auto-correlation function, $C_{VV}$, in the inset, show the loss of directionality in swimming. \plb{f} Distribution of propulsion speeds $V$ for increasing $Pe$, with mean and standard deviation of speeds in the inset. See also Supporting Videos S2-S5. The color bar relating experimental $Pe$ estimates to the viscosity ratio  $\mu=\mu^o/\mu^i$ applies to all subsequent figures.
}
\end{figure*}

\section*{Destabilised motion with increasing P\'eclet number}
We begin, however, with an overview of the droplet dynamics using trajectory plots and statistical analyses of speed and orientational persistence taken from bright-field microscopy (\figref{fig:statistics}).
With increasing $Pe$, the droplet propulsion changes from uniform speeds and persistent motion to unsteady motion with
abrupt reorientations (\figref[a-d]{fig:statistics}).
We define  $P(|\delta\theta(t)|)$ as the distribution of the reorientation angle $\delta\theta$ of the 2D droplet velocity $\vec{V}(t)$ during a fixed time step $\delta t$~\cite{bhattacharjee2019_bacterial}, \begin{equation}\delta\theta(t) = \arctan\left(\frac{\vec{V}(t)\times \vec{V}(t+\delta t)}{\vec{V}(t) \cdot \vec{V}(t+\delta t)}\right).\label{eqn:dtheta}\end{equation}

$P(|\delta\theta(t)|)$ broadens significantly, corresponding to more frequent and sharper reorientation events~(\figref[e]{fig:statistics}). 
The faster decay of the angular velocity autocorrelation function, 
\begin{equation}C_{VV}(t)=\left\langle\frac{\vec{V}(t_0 + t)\cdot\vec{V}(t_0)}{|\vec{V}(t_0 + t)| |\vec{V}(t_0)|}\right\rangle_{t_0},\label{eqn:cvv}\end{equation}
illustrates the loss of directionality with increasing $Pe$ (\figref[e]{fig:statistics}, inset).
\figref[f]{fig:statistics} shows that at sufficiently large $Pe$, the speed distribution $P(V)$ includes values as small as zero (stopping events) and, surprisingly, as large as 70\,$\mu$m/s, much greater than the uniform speed of $30\,\mu$m/s observed for low $Pe\approx 4$.
While the mean speed barely changes with $Pe$, the standard deviation of $V$ grows by over one order of magnitude (\figref[f]{fig:statistics}, inset).
Hence, both the rotational and the translational motion of the swimmer are destabilised with increasing $Pe$, similar to recent numerical studies of solid phoretic particles~\cite{hu2019_chaotic}.
Note that the thermal fluctuations ($O(k_bT/2 R_d)\sim 10^{-16}$\,N) are negligible compared to the hydrodynamic drag force ($O(6\pi \mu^o R_d V)\,\gtrsim\, 10^{-10}$\,N), such that thermal noise is an unlikely cause for the unsteady swimming. 

\begin{figure*}
\includegraphics[width=\textwidth]{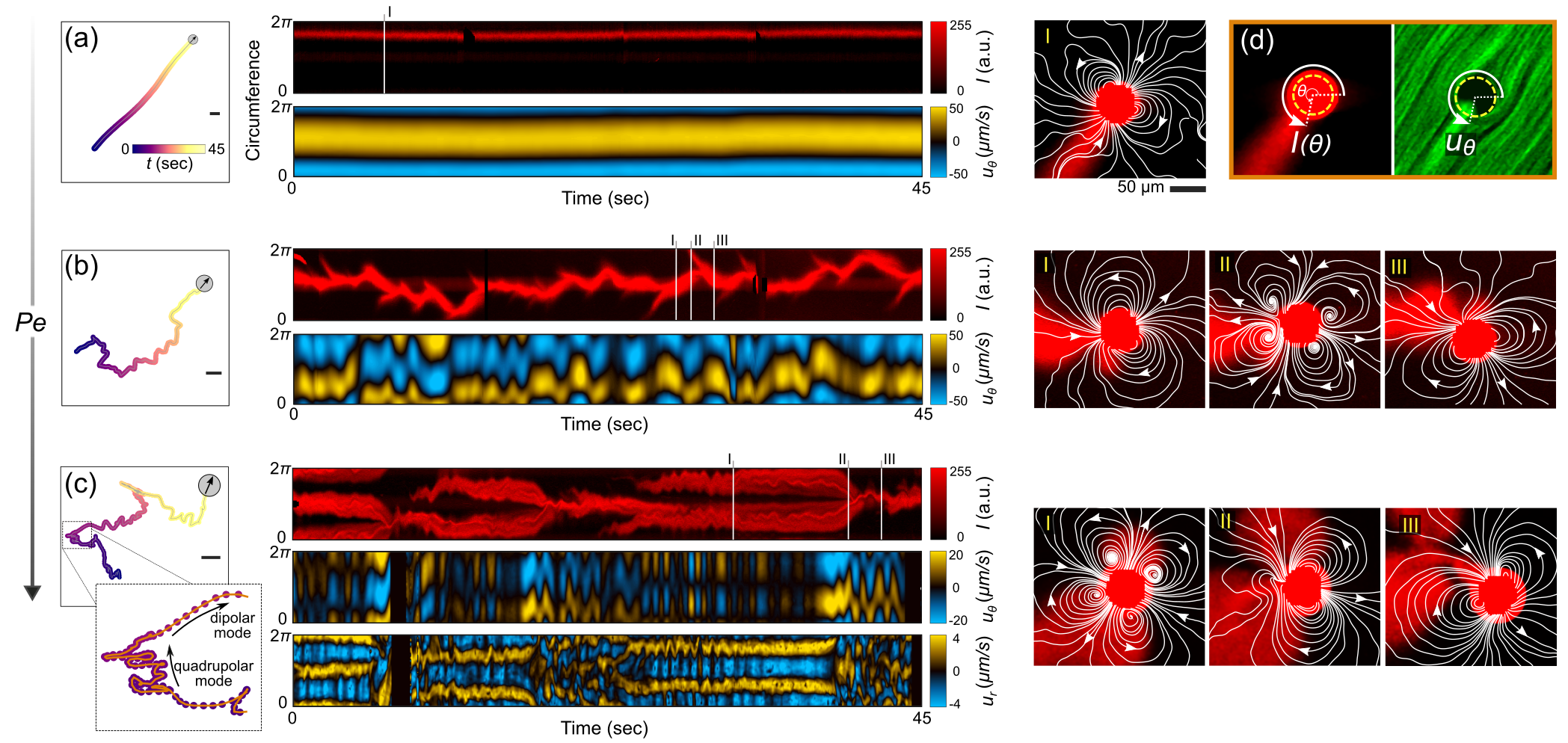}%
\caption{\label{fig:kymographs} Signatures of unsteady dynamics in the time evolution of chemical and hydrodynamic fields.
Rows \plb{a}, \plb{b} and \plb{c} correspond to $Pe\approx4$, $Pe\approx36$ and $Pe\approx293$, respectively; Left column, trajectories colour coded by time; middle column, kymographs of $I$ and $u_{\theta}$ during 45 seconds of propulsion; right column, selected red channel images, overlaid by the flow streamlines in the laboratory reference frame. Each frame corresponds to the point in time indicated on the kymographs by I, II or III.
Panel \plb{d} defines the mapping of the profiles of red light intensity $I$ (filled micelle concentration) and tangential velocity, $u_{\theta}$, around the droplet circumference onto the $y$ axis of the kymographs in the middle column.
All $u_{\theta}$ profiles are in the translational droplet reference frame, but with $\theta=0$ fixed at the laboratory $x$ direction to visualize the reorientation dynamics. In \plb{c}, the third kymograph corresponds to the radial velocity $u_r$ in the laboratory reference frame to better depict the quadrupolar symmetry of the flow field. 
The second hydrodynamic mode starts to appear at intermediate Pe and dominates the dynamics for high Pe.
See also Appendix \ref{appsec:figures}, \figref{sifig:ComplementaryFig4} for additional flow field examples. All scale bars are 50\,$\mu$m. 
\protect}
\end{figure*}

\section*{Signatures of unsteady dynamics in the time evolution of chemical and hydrodynamic fields}
To investigate the origin of this unsteady behaviour, we studied the evolution of chemical and hydrodynamic fields around the droplet. We extracted the tangential flow velocity $u_\theta(\theta)$ and the red fluorescence intensity $I(\theta)$ of the chemical field close to the interface (\figref[d]{fig:kymographs}, Appendix \ref{sisec:improc}), and mapped them in kymographs $I(\theta, t)$ and $u_\theta(\theta, t)$.

For low $Pe\approx 4$, at persistent propulsion, $I(\theta,t)$ shows a single fixed-orientation band marking the origin of the filled micelle trail at the rear stagnation point of the droplet (\figref[a]{fig:kymographs} and Video S6). 
The two bands in $u_\theta(\theta,t)$ correspond to a steady flow field with dipolar symmetry that is consistent with the $I(\theta,t)$ profile. On the right side of \figref[a]{fig:kymographs} we have superimposed the streamlines of this dipolar flow field on the corresponding chemical micrograph at the time marked by I in the $I(\theta,t)$ kymograph.

For intermediate $Pe \approx 36$ (\figref[b]{fig:kymographs}, Video S7), $I(\theta,t)$ shows secondary branches forming at the anterior stagnation point of the droplet and subsequently merging with the main filled micelle trail. 
This coincides with a transient second hydrodynamic mode with quadrupolar symmetry (\figref[b,II]{fig:kymographs}), causing the accumulation of an additional aggregate of filled micelles at the droplet anterior (see also Appendix \ref{appsec:figures},  \figref{sifig:ComplementaryFig4} for additional flow field examples). 

The ratio of the diffusive $(R_d^2/D_\text{{\it fm}})$ to advective $(R_d/V)$ time scales for the migration of filled micelles is $\frac{VR_d}{D_\text{{\it fm}}}\gg1$ for all experiments, assuming a diffusion coefficient $D_\text{{\it fm}}=k_BT/6\pi\mu^or_\text{{\it fm}}$, with a micellar radius of $O(r_\text{{\it fm}})\sim 2.5$\,nm. 
Therefore, the aggregate is unlikely to dissipate by diffusion, and will continue to grow as long as the quadrupolar mode exists.
However, this mode is not stable. Eventually, the dipolar mode dominates and advects the secondary aggregate towards the main trail (\figref[b,III]{fig:kymographs}). 
The transport of the aggregate along one side of the droplet locally disturbs the interfacial flow, leading to an abrupt reorientation of the swimming direction (\figref[b,I-III]{fig:kymographs}).
 As shown in the trajectories in \figref[b and c]{fig:statistics}, these reorientation events become more frequent with increasing $Pe$; accordingly, $u_\theta$ in \figref[b]{fig:kymographs} exhibits quasi-periodic reorientation patterns.
 
For high $Pe\approx 293$ (\figref[c]{fig:kymographs}, Video S8), the quadrupolar mode eventually prevails, resulting in a predominantly symmetric extensile flow around the droplet (\figref[c,I]{fig:kymographs}), as shown by a pronounced fourfold pattern in the additional kymograph $u_r(\theta,t)$ of the radial velocity. Due to the non-propelling quadrupolar mode the droplet is trapped in place.  
The gradual accumulation of filled micelles at the two stagnation points with radially outward flow manifests in two stable branches in the chemical kymograph (marked by I in \figref[c]{fig:kymographs}).
 The growth of the two micellar aggregates locally generates a lateral chemical gradient, which eventually pushes the droplet out of its self-made trap.
Concomitantly, the two points of filled micelle emission move along the droplet interface and merge on the new rear side of the droplet into a single filled micelle trail (\figref[c,II and III]{fig:kymographs}). 
The chemorepulsion from the local field micelle gradient induces an apparent dipolar mode which gradually decays as the droplet leaves the self-made trap.
Now, the quadrupolar mode re-saturates, with an aggregate growing at the droplet anterior, until the droplet is trapped again and a new bimodal `stop-and-go' cycle begins. Since the escape direction is always lateral, consecutive runs are approximately perpendicular, resulting in the sharp reorientation events apparent in the trajectories in \figref[c]{fig:kymographs} and \figref[d]{fig:statistics}, as well as the broadening $|\delta\theta|$ distribution in \figref[e]{fig:statistics}.
\begin{figure*}
\centering\includegraphics[width=.9\textwidth]{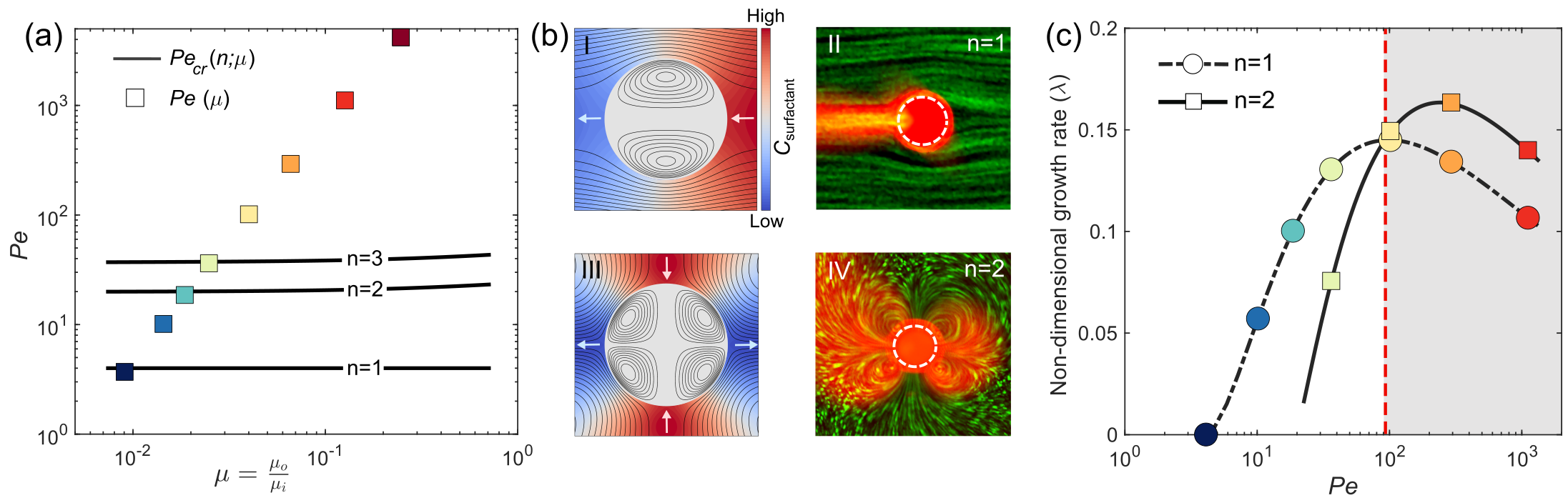}%
\caption{\label{fig:modestab} Dependence of hydrodynamic modes on the P\'eclet number. \plb{a} Critical P\'eclet $Pe_{cr}$ values (black lines), necessary for the onset of different hydrodynamic modes $(n)$, with varying $\mu$; The markers ($\square$) show the P\'eclet number $Pe$ (Eq.~\ref{eqn:pe}) which increases with $\mu$. The colour code is taken from \figref{fig:statistics} \plb{b} (top) Steady self-propulsion of the active droplet; theoretical solution for $n=1$ mode (left) and experimental streak image for low $Pe$ (right); (bottom) the extensile flow corresponding to $n=2$ mode (left) and the experimental image for higher $Pe$ (right). The theoretical and the experimental flow fields are in the swimmer reference frame. \plb{c} Instability growth rates corresponding to the first two hydrodynamic modes as a function of $Pe$. Beyond the dashed vertical line (grey region) $\lambda_{n=2}>\lambda_{n=1}$ and thus the $n=2$ mode is dominant.}
\end{figure*}
\section*{Dependence of hydrodynamic modes on the P\'eclet number}
In order to understand the dependence of the onset of bimodal motility on $Pe$, we analysed the underlying advection-diffusion problem for the active droplet within the framework of an axisymmetric Stokes flow as established in refs. ~\cite{michelin2013_spontaneous,morozov2019_nonlinear, morozov2019_self-propulsion, leal2007_advanced} (see \figref{fig:modestab}, and Appendix \ref{sisec:hdmodel}).
At the smallest value of $\mu$, $Pe$ is approximately equal to the critical value of 4 necessary for the onset of the first hydrodynamic mode ($n=1$), i.e. the mode with dipolar flow symmetry~\cite{michelin2013_spontaneous,morozov2019_nonlinear, morozov2019_self-propulsion}. 
With increasing $\mu$, $Pe$ (markers in \figref[a]{fig:modestab}) eventually exceeds the critical values necessary for the onset of the higher hydrodynamic modes (lines in \figref[a]{fig:modestab}), specifically the second hydrodynamic mode ($n=2$), i.e. the mode with quadrupolar symmetry.
A linear stability analysis around an isotropic, quiescent base state (see Appendix \ref{sisec:linstab} and~\cite{michelin2013_spontaneous, morozov2019_self-propulsion}), which is the idealized starting point for each experiment, shows that for small to moderate $Pe$, the non-dimensionalised instability growth rate $\lambda$ for $n=1$ exceeds that for $n=2$ (\figref[c]{fig:modestab}).
Accordingly, for lower $Pe$, $n=1$ dominates, resulting in steady self-propulsion stemming from the fore-aft asymmetry of the surfactant distribution (\figref[b,I]{fig:modestab}). 
Consequently, the active droplet exhibits persistent steady translation (trajectories in \figref[a,b]{fig:statistics}) with a dominant dipolar flow field (\figref[b,II]{fig:modestab} and \figref[a]{fig:kymographs}).
However, for $Pe\gtrsim92$, $n=2$ (\figref[b,III]{fig:modestab}) has a faster instability growth rate (\figref[c]{fig:modestab}), thereby becoming the dominant mode  when evolving from the quiescent base state.
Accordingly, the droplet is initially stuck in a non-propelling mode with a quadrupolar flow field (similar to \figref[b,IV]{fig:modestab}). 
Such quadrupolar flow field gives rise to the filled micelle field with the two points of outflux. 
The synergy between the $n=2$ mode and the transiently-growing filled micelle field subsequently results in the onset of the bimodal `stop-and-go' motion of the droplet for moderate to higher $Pe$ (trajectories in \figref[c,d]{fig:statistics}).
Since we observe in experiments with $Pe\gtrsim 100$ that the active droplet experiences sustained periods of dynamical arrest during which it remains stationary with a surrounding extensile flow (\figref[c]{fig:kymographs}), it appears that the $n=2$ mode can also evolve from a non-quiescent state and prevail in a similar P\'eclet regime as derived from the performed  stability analysis.
Note that we restrict our analysis to the first two hydrodynamic modes since these two are solely responsible for the droplet propulsion and the associated far-field hydrodynamic disturbance.

\begin{figure*}
\centering\includegraphics[width=0.8\textwidth]{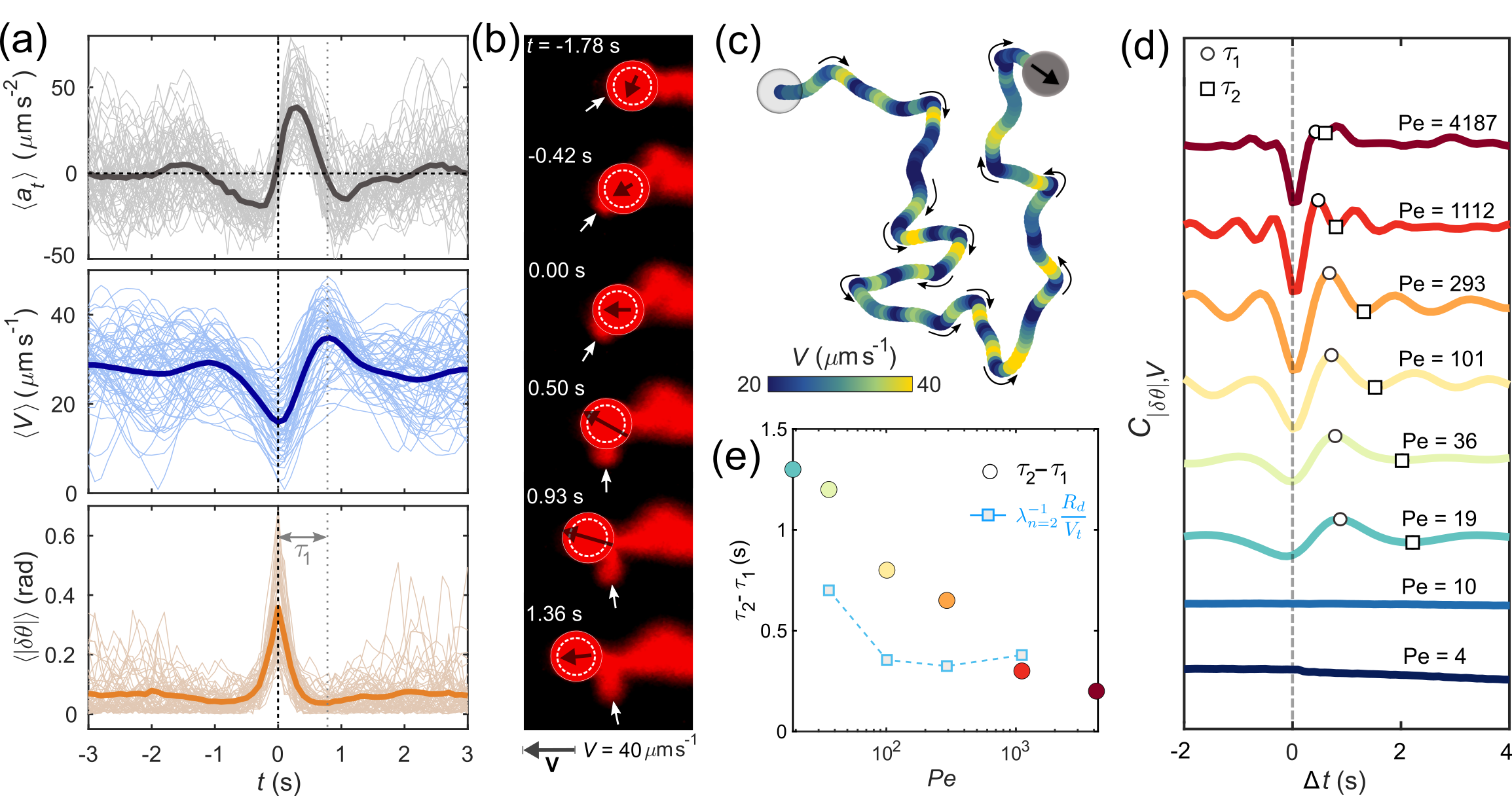}
\caption{\label{fig:correlation} Interactions with self-generated chemical gradients cause speed bursts at reorientation events.
\plb{a} Conditional averaging of tangential acceleration, $a_t$, speed, $V$, and reorientation angle, $|\delta\theta|$, for abrupt reorientation events at $Pe\approx36$ (see Appendix \ref{appsec:figures}, \figref{sifig:Cutoff}, for an illustration of the identification criteria). The dotted line marks the maximum speed at $t=\tau_1$ after reorientation.
\plb{b} Video stills of the chemical field for one such event with $t = 0\,s$ set to the point of minimum speed; white arrows track the accumulation of the secondary filled micelle aggregate at the anterior stagnation point and its advection along the interface, black arrows correspond to the droplet velocity vector. The droplet speed is maximal when the secondary aggregate and the trail merge at $t = 0.93 s$. See also Videos S9 and S10.
 \plb{c} An example trajectory for $Pe\approx36$. Any reorientation event (curved arrows) is preceded by a deceleration and followed by an acceleration. The lowest speed occurs at the point with the highest curvature.
 \plb{d} Correlation function between reorientation angle and speed, $C_{|\delta\theta|,V}(\Delta t)$ for increasing $Pe$.  Times $\tau_1$ and $\tau_2$ (next reorientation event) are identified by the respective peak and dip in $C_{|\delta\theta|,V}$. 
\plb{e} Time scale for the growth of the $n=2$ mode vs. corresponding $Pe$: experimentally obtained, $\tau_2-\tau_1$ ($\circ$), compared to values from stability analysis, $\lambda^{-1}_{n=2}R_d/V_t$ ($\Box$).}
\end{figure*}

\section*{Interactions with self-generated chemical gradients cause speed bursts}
It remains to explain the broadening of $P(V)$ with increasing $Pe$ (\figref[e]{fig:statistics}), particularly the remarkable bursts in speed for high $Pe$. 
While the dipolar mode is propulsive, the quadrupolar mode is not. Hence, the growth and decay of the respective modes will affect the droplet speed. 
As shown in \figref{fig:kymographs}, recurrent transitions between the two hydrodynamic modes lead to abrupt reorientation events; we therefore investigated the correlation between changes in speed and reorientation angle $|\delta \theta|$.

In a typical trajectory for intermediate $Pe\approx36$, each sharp turn is preceded by a deceleration and followed by an acceleration, as shown in the plot of the positional data colour-coded by speed in \figref[c]{fig:correlation}.
Signatures of these correlations in the droplet dynamics appear in the conditional averages 
\begin{equation}\left\langle X(t-t_i)\right\rangle_i \text{ if } |\delta\theta(t_i)|>0.2;\frac{{\rm d}(\delta\theta (t_i))}{{\rm d} t} =0\end{equation}%
of $|\delta\theta|$, $V$ and tangential acceleration $a_t$ as quantities $X$ for all sharp reorientation events $i$ in the trajectory, centered at $t=t_i$ of maximum $|\delta \theta|$~(\figref[a]{fig:correlation}); the events were identified by choosing a threshold value of $|\delta\theta|>0.2$ (see Appendix \ref{appsec:figures}, \figref{sifig:Cutoff}).

We can now directly compare these dynamics to the higher resolution fluorescence data taken at $Pe\approx36$ presented in the kymographs in \figref[b]{fig:kymographs}. 
\figref[b]{fig:correlation} shows a series of micrographs of the chemical field, with arrows marking the droplet velocity vector (black) and the position of the secondary filled micelle aggregate (white).
The aggregate accumulates, is then entrained and finally merges with the posterior trail, corresponding to the creation and merging of a secondary chemical branch in the kymograph. 

For $t<0$ the droplet decelerates while the secondary aggregate is accumulating. 
$t=0$ marks the point in time where $V$ is minimal and the aggregate is on the cusp of leaving the anterior stagnation point.
For $t>0$, the aggregate is advected to the droplet posterior and the droplet accelerates due to the re-saturation of the dipolar mode.
$V$ peaks once the aggregate has merged with the main trail --- creating an amplified fore-aft gradient --- at $t\approx 1\,$s, which is comparable to the advective timescale $R_d/V\approx1\,$s. 
In the wide-field data analysis in \figref[a]{fig:correlation}, this is the time $\tau_1$ it takes the droplet to reach maximum speed after a reorientation.

We now use the correlation function between $V$ and $|\delta\theta|$, $C_{|\delta\theta|,V}(\Delta t)=\big \langle {|\delta\theta(t)|\cdot V(t+\Delta t)}\big\rangle_{t}$, plotted in \figref[d]{fig:correlation}, to estimate the growth times of the second mode from our data for $Pe>10$. 
Since $V$ is minimal at maximum $|\delta\theta(t)|$ (\figref[d]{fig:correlation}), $C_{|\delta\theta|,V}(\Delta t)$ dips at $\Delta t=0$. It subsequently peaks at the point of maximum $V$ with a time delay $\Delta t=\tau_1$, when the contribution of the propulsive dipolar flow is maximal. 
The next dip at a time $\tau_2>\tau_1$ marks the next reorientation event; based on the discussion pertaining to \figref{fig:kymographs} and \figref[c]{fig:modestab}, for moderate to high $Pe$, $\tau_2-\tau_1$ approximately corresponds to the time scale for the growth and re-saturation of the $n=2$ mode during the bimodal motility (i.e. starting from a non-quiescent base state). 
Nevertheless, we compare this experimentally obtained $\tau_2-\tau_1$ with the theoretical growth times for the $n=2$ mode starting from the isotropic base state, $\lambda^{-1}_{n=2}R_d/V_t$ (\figref[c]{fig:modestab}), for different values of $Pe$. 
\figref[e]{fig:correlation} shows that these two time scales, which are strictly speaking different, still are of the same order of magnitude and show similar decreasing trend with increasing $Pe$. 
We note that the growth time of the dipolar flow above $Pe\approx100$ cannot be used for comparison to  $\lambda_{n=1}$, since this flow is imposed by the lateral chemical gradient. 
However, we can assume that this gradient increases with $Pe$, resulting in faster acceleration, markedly higher swimming speeds, and hence, reduced $\tau_1$, as observed experimentally (\figref[d]{fig:correlation}).

\begin{figure}
\includegraphics[width=\columnwidth]{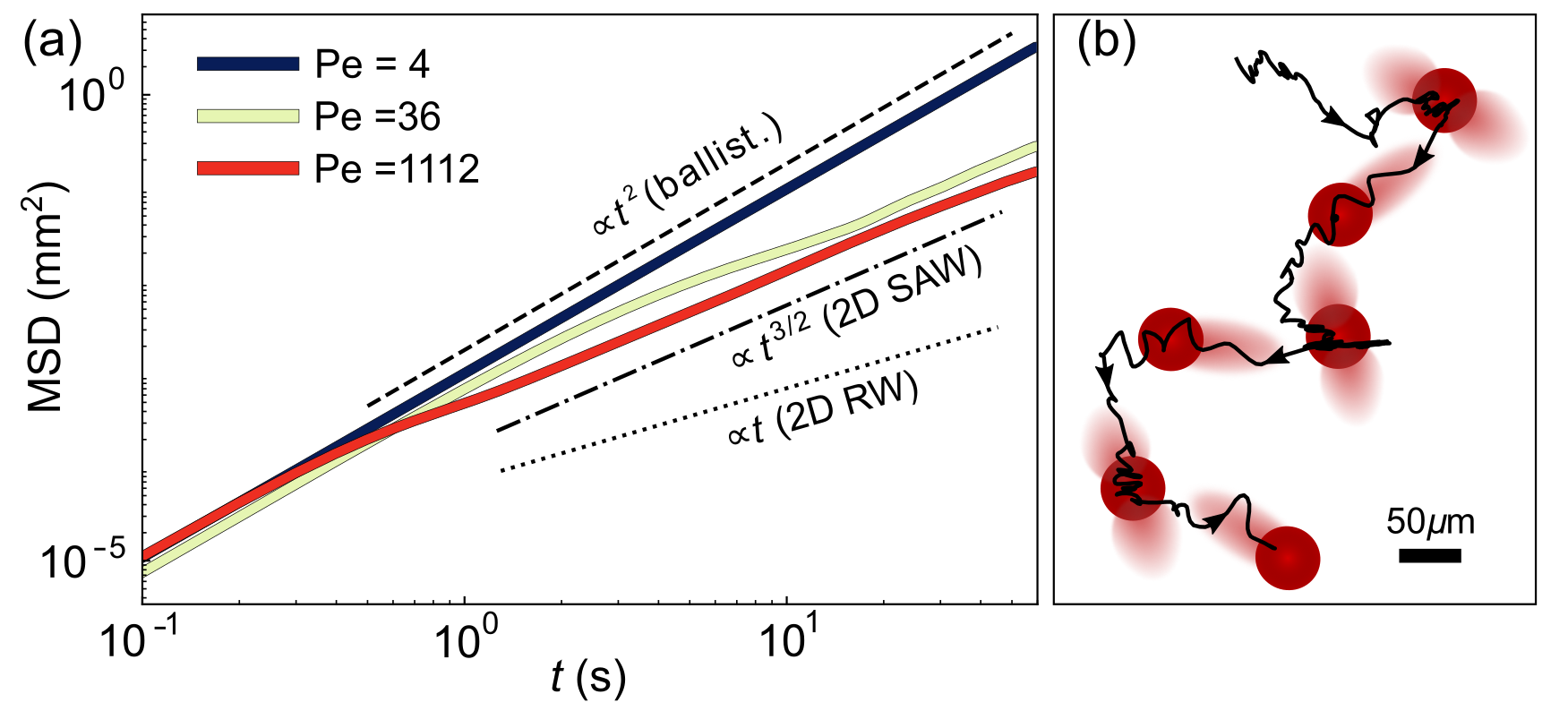}
\caption{\label{fig:SAW} Anomalous diffusive swimming. \plb{a} Mean squared displacement profiles  of experimental trajectories for different $Pe$. Dashed lines mark the predicted scaling for ballistic motion, $\propto t^2$, 2D self-avoiding walk (SAW), $\propto t^{3/2}$, and random walk (RW), $\propto t$. For higher $Pe$, there is a transition from ballistic to 2D SAW. \plb{b} A segment of the trajectory associated with the SAW and schematics of the droplet exhibiting bimodal swimming causing the SAW. See also Appendix \ref{appsec:figures}, \figref{sifig:BimodalSignal}.}
\end{figure}

\section*{Consequences for spatial exploration}
Reminiscent of gait switching dynamics in biological locomotion, we have demonstrated the emergence of complex swimming behaviour in a minimal active droplet system by tuning the P\'eclet number. 
We found a transition from persistent swimming at low $Pe$ to chaotic bimodal swimming at high $Pe$ --- the latter results
from the excitation of higher hydrodynamic modes beyond critical $Pe$ values, while the continuous switching between them is caused by the self-generated chemical gradient in the environment.

This gradient sensitivity causes trail avoidance~\cite{jin2017_chemotaxis}, which in turn affects the way these droplet swimmers explore their environment.
With increasing reorientation frequency, we find a transition from quasi-ballistic propulsion to a 2D self-avoiding walk (2D SAW). This effect is illustrated by the trajectories in \figref[a-d]{fig:statistics}, and also by the fact that $C_{VV}$ in \figref[e]{fig:statistics} does not decay to zero. For a statistical analysis we have plotted mean squared displacements for selected $Pe$ values in \figref[a]{fig:SAW}, which reproduce the expected scaling with $t^2$ (ballistic) for $Pe\approx4$ and a transition to $t^{3/2}$ (2D SAW,~\cite{slade1994_self-avoiding}) for $Pe\gtrsim36$, with the crossover time decreasing with increasing $Pe$. 
While transitions to random walks governed by run-and-tumble gait switching are common in bioswimmers \cite{najafi2018_flagellar}, self-avoidance requires chemical self-interaction~\cite{golestanian2009_anomalous}. 

Examples of anomalous diffusion driven by repulsive biochemical signalling have been found in the spreading of slime molds~\cite{reid2012_slime,cherstvy2018_non-gaussianity} --- active droplets can show analogous behaviour based on purely physicochemical mechanisms. 
\section*{Conclusion}
In this work, we demonstrated that the manner in which hydrodynamic and self-generated chemical fields are coupled determines the nonlinear dynamics of autophoretic micro-swimmers. 
The fluorescence-based visualisation technique used to simultaneously probe this coupling can provide insight into many recent autophoretic models~\cite{hokmabad2019_topological,izzet2020_tunable,maass2016_swimming,izri2014_self-propulsion, michelin2013_spontaneous,schmitt2013_swimming,meredith2020_predator-prey}.
For example, extensive theoretical studies~\cite{jabbarzadeh2018_viscous,nasouri2020_exact,lippera2020_bouncing,yang2019_autophoresis} have demonstrated the importance of quantifying far-field and near-field contributions, coupling to chemical fields and the effects of confinement to understand how swimmers approach each other or form bound states, which is vital to nutrient entrainment, food uptake and mating in bioswimmers. 

While many micro-swimmer models incorporate unsteady dynamics via stochastic fluctuations, we have shown that the interplay of nonlinear dynamics and interaction with the history of motion also allows for the emergence of memory-driven chaotic behaviour.
An appealing example from a different field are droplet walkers on a vibrated bath~\cite{couder2005_walking}, which show a transition from persistent to a bimodal, stop-and-go motion based on an effective `system memory' parameter~\cite{hubert2019_tunable,valani2019_superwalking}. 
The corresponding theoretical framework~\cite{hubert2019_tunable} is general enough to also apply to bimodal chaotic motion in droplet swimmers.

We acknowledge fruitful discussions with Stephan Herminghaus, Arnold Mathijssen and Prashanth Ramesh, as well as financial and organisational support from the DFG SPP1726 ``Microswimmers'' (CCM, RD, BVH), the ERC-Advanced Grant ``DDD''  (DL, MJ), and the Max Planck Center for Complex Fluid Dynamics.


\cleardoublepage

\appendix
\counterwithin{figure}{section}

\section{Materials and Methodology}\label{appsec:materials}

\subsection{Materials and characterisation}
Our samples consisted of droplets of (S)-4-Cyano-4'-(2-methylbutyl)biphenyl  (CB15) doped with the fluorescent dye Nile Red in an aqueous solution of the cationic surfactant tetradecyltrimethylammonium bromide (TTAB) corresponding to 5\,wt.\%  (50 mg in 1 ml of solution) in pure water, with a critical micelle concentration of \textit{CMC}$ = 0.13$\,wt.\%. 
We purchased CB15, TTAB, and Nile Red from commercial suppliers (Synthon Chemicals and Sigma-Aldrich) and used them as is. 
We controlled the viscosity of the swimming medium, $\mu_o$, by adding glycerol to the aqueous TTAB solution. 

\begin{figure}
\includegraphics[width=0.9\columnwidth]{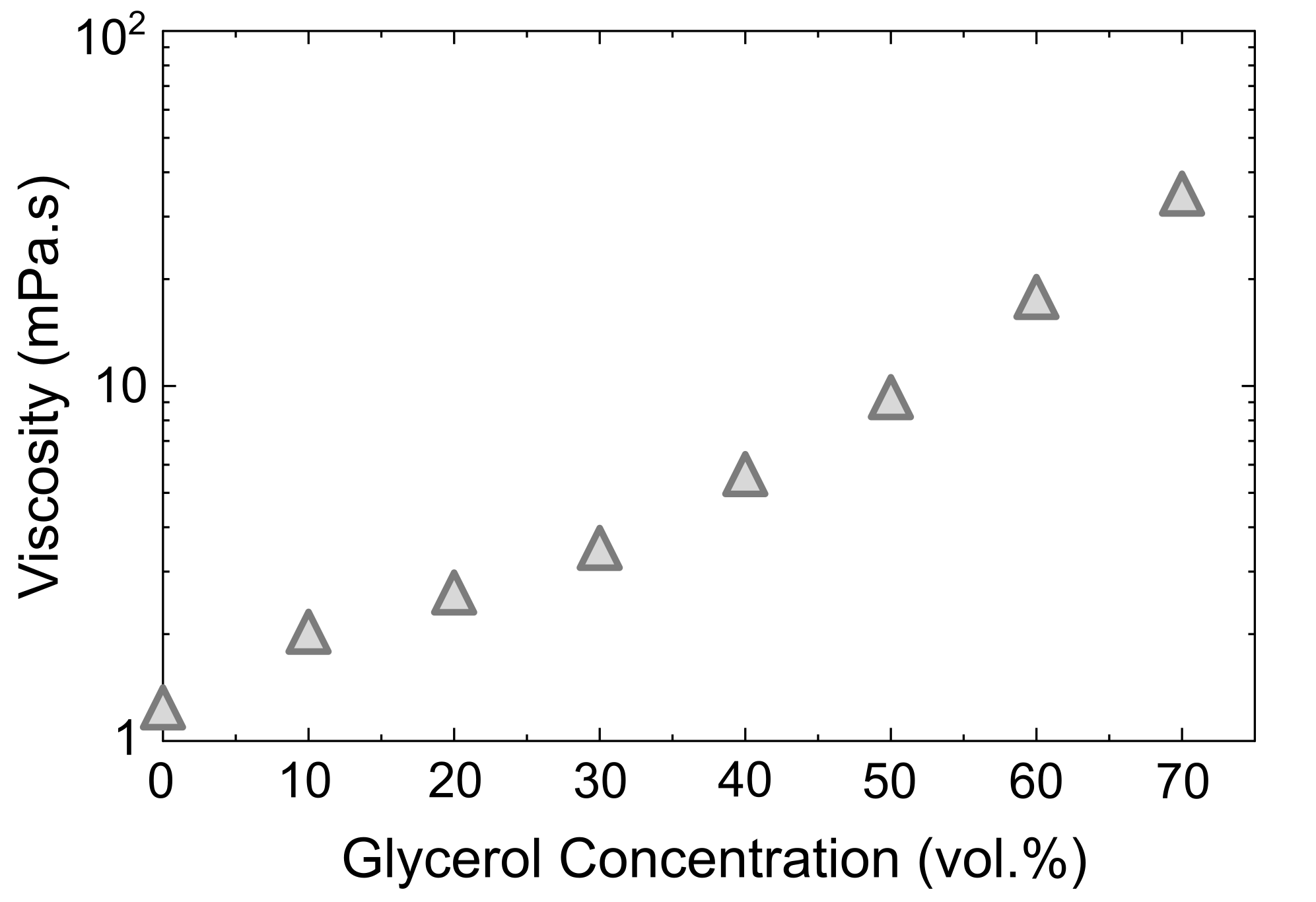}
\caption{\label{sifig:Viscosity} Viscosity of the swimming medium, a mixture of water, glycerol and TTAB surfactant, for increasing glycerol/water ratios. Surfactant concentration is 50\,mg in 1\,ml of solution.}
\end{figure}

We used an Anton Paar MCR 502 rotational rheometer to characterise the shear viscosity of water-glycerol-surfactant solutions (\figref{sifig:Viscosity}). 
Experiments were carried out using a cone-plate geometry, to find shear-rate versus shear-stress curves at a fixed temperature, and viscosity versus temperature a fixed shear rate. 
To limit effects of solution evaporation, the cone-plate geometry was surrounded by a water bath and covered by a Peltier hood. 
Over the shear rate range $0.01\,\rm s^{-1} <\dot{\gamma}<100\,\rm s^{-1}$, viscosity was found to be constant, such that our solutions are well-described as Newtonian, as should be expected: Water/glycerol mixtures are used as Newtonian standard media throughout the existing literature.

\begin{figure}
\includegraphics[width=0.9\columnwidth]{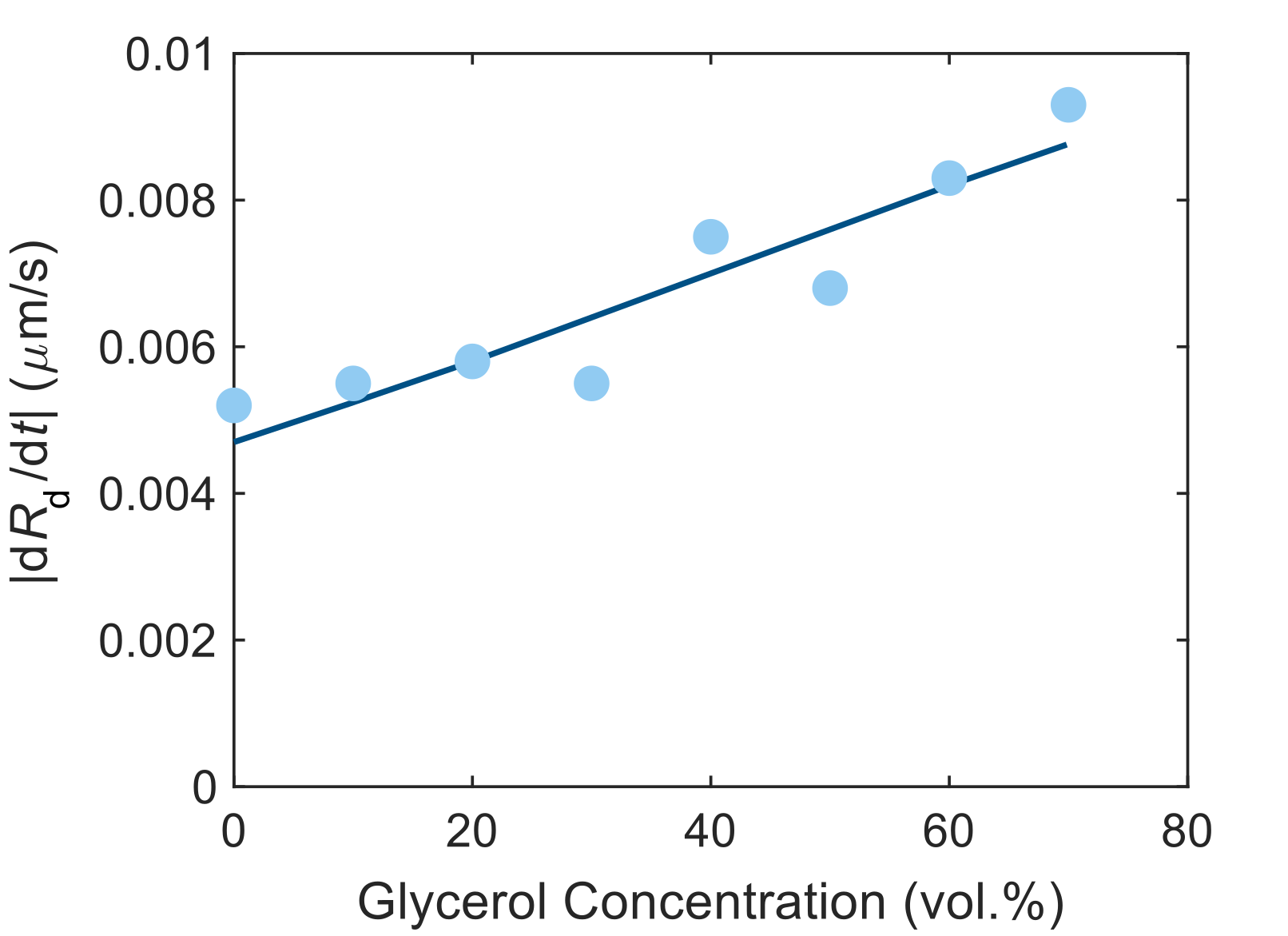}
\caption{\label{sifig:Solubilisation} Solubilisation rate $|{\rm d}R_d/{\rm d}t|$, for increasing glycerol/water ratios, blue line marks a linear regression fit to $y=0.000058x+0.0047$. Surfactant concentration is 50\,mg in 1\,ml of solution.}
\end{figure}

To estimate the surfactant consumption rate $q_s$ in \eqref{eqn:pe}, we extracted the droplet shrinking rate ${\rm d}R_d/{\rm d}t$ from the bright field microvideography data presented in \figref{fig:statistics}. 
We found a moderate dependence on the glycerol fraction (\figref{sifig:Solubilisation}), which we included as a first order approximation, via linear regression (blue line),  to evaluate $q_s$  in the $Pe$ estimates in the main manuscript.

\subsection{PDMS soft lithography for droplet generation}\label{sisec:lithography}
For the production of monodisperse oil droplets, we fabricated microfluidic channels in-house, using standard soft lithography techniques. 
First, 2D photomasks were designed in AutoCad, and then printed onto an emulsion film in high-resolution (128,000\,dpi) by a commercial supplier (JD Photo-Tools). Next, the photoresist SU-8 3025 (MicroChem) was spin-coated onto a 4 inch diameter silicon wafer (Si-Mat), where spin-speed and duration were adjusted to give a controllable uniform thickness. 
A negative mold was cured in the SU-8 through the photomask by UV light exposure. 
After further chemical treatment with photoresist developer, uncured SU-8 was removed, leaving behind cured SU-8 microstructures on the silicon wafer.

\begin{figure}
\includegraphics[width=\columnwidth]{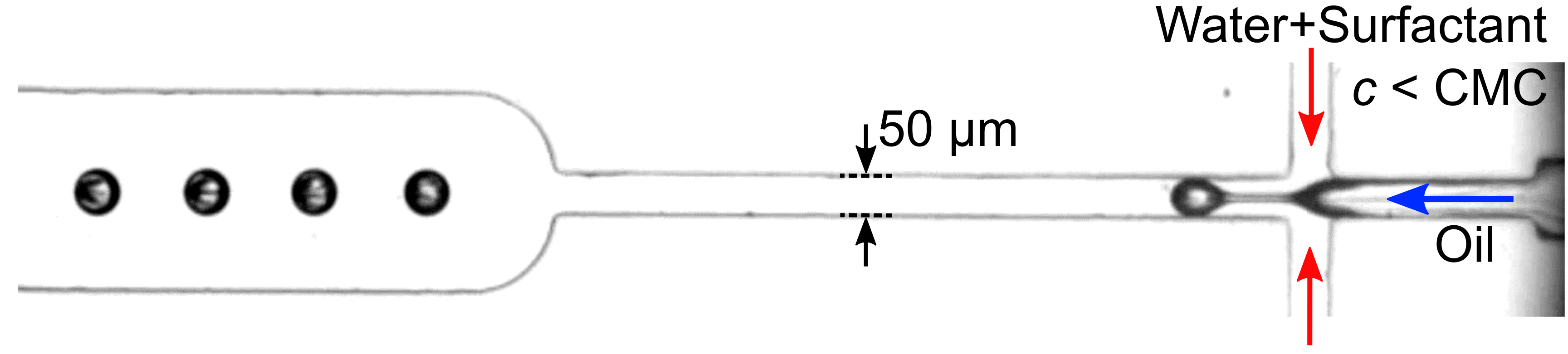}
\caption{The microfluidic chip used to produce monodisperse oil droplets in a surfactant solution.}\label{sifig:Microfluidics}
\end{figure}

We then poured a poly(dimethyl siloxane) (PDMS, Sylgard 184, Dow Corning) mixture of 10:1 volumetric ratio of base to cross-linker over the wafer, and baked for 2 hours at 80$\,^\circ$C, producing a solid PDMS layer with microstructured indentations. 
We peeled the indented PDMS from the wafer, and punched holes through it to create liquid inlets/outlets at opposing ends of the channels. 
The structured PDMS surface, as well as a glass coverslip, were cleaned and treated with partial pressure air-plasma (Pico P100-8; Diener Electronic GmbH + Co. KG) for 30 seconds, and then pressed together, bonding the two surfaces.
\figref{sifig:Microfluidics} shows a micrograph of such a PDMS chip during droplet production.

The walls of these microfluidic chips were selectively treated to hydrophilise the channels where surfactant solution will flow. This prevents oil from wetting the walls during droplet production. 
We followed the technique of Petit et al.\cite{petit2016_vesicles-on-a-chip}: First, the channel walls were oxidised by a 1:1 mixture of hydrogen peroxide solution (H2O2 at $30 \,\rm wt.\%$ , Sigma-Aldrich) and hydrochloric acid (HCl at $37 \,\rm wt.\%$, Sigma-Aldrich). This mixture was flushed through the channels for approximately 2 minutes by using a vacuum pump system. After the oxidation, the channel was rinsed by flushing double distilled water for 30 seconds. 
Next, a $5 \,\rm wt.\%$ solution of the positive polyelectrolyte poly(diallyldimethylammonium chloride) (PDADMAC, Sigma-Aldrich) was flushed for 2 minutes through the oxidised channel of the device. The PDADMAC binds to the activated channel walls by ionic interactions. 
Finally, a $2 \,\rm wt.\%$ solution of the negative polyelectrolyte poly(sodium 4- styrenesulfonate) (PSS, Sigma-Aldrich) was flushed for 2 minutes.

\subsection{Droplet generation}

Once the chips had been treated, we mounted syringes of oil and $0.1 \,\rm wt.\%$ aqueous TTAB solution to a microprecision syringe pump (NEM-B101-02B; Cetoni GmbH), and connected these to the two inlets of the microfluidic chip via Teflon tubing (39241; Novodirect GmbH), and tuned the flow speed through the chip until the desired droplet size was reached. Once droplet production was monodisperse (after approximately 5 minutes) and at a steady state, these droplets were collected in a bath of $0.1 \,\rm wt.\%$ TTAB solution. This solution is of a high enough concentration to stabilize the droplets against coalescence, but not high enough to induce solubilization. 

\subsection{Fabrication of the observation Hele-Shaw cell}
The swimming behaviour of the droplets was observed in a quasi-2D Hele-Shaw reservoir, which we fabricated directly from SU-8 photoresist without PDMS casting. To fabricate the reservoirs we therefore used a photo-mask with inverted polarity. We spin-coated the photoresist directly onto a glass slide ($50\times75\, \rm mm^2$) and followed the same procedure for photo-lithography as outlined in section~\ref{sisec:lithography}. This resulted in a layer of crosslinked SU-8 (thickness $\approx 60\,\mu$m) with reservoirs of the dimensions $8\times13\,$mm. These reservoirs were filled with the samples, sealed with a glass cover slip and put under a microscope.

\subsection{Double-channel fluorescent microscopy technique}\label{sisec:DCFM}
\begin{figure}
\includegraphics[width=\columnwidth]{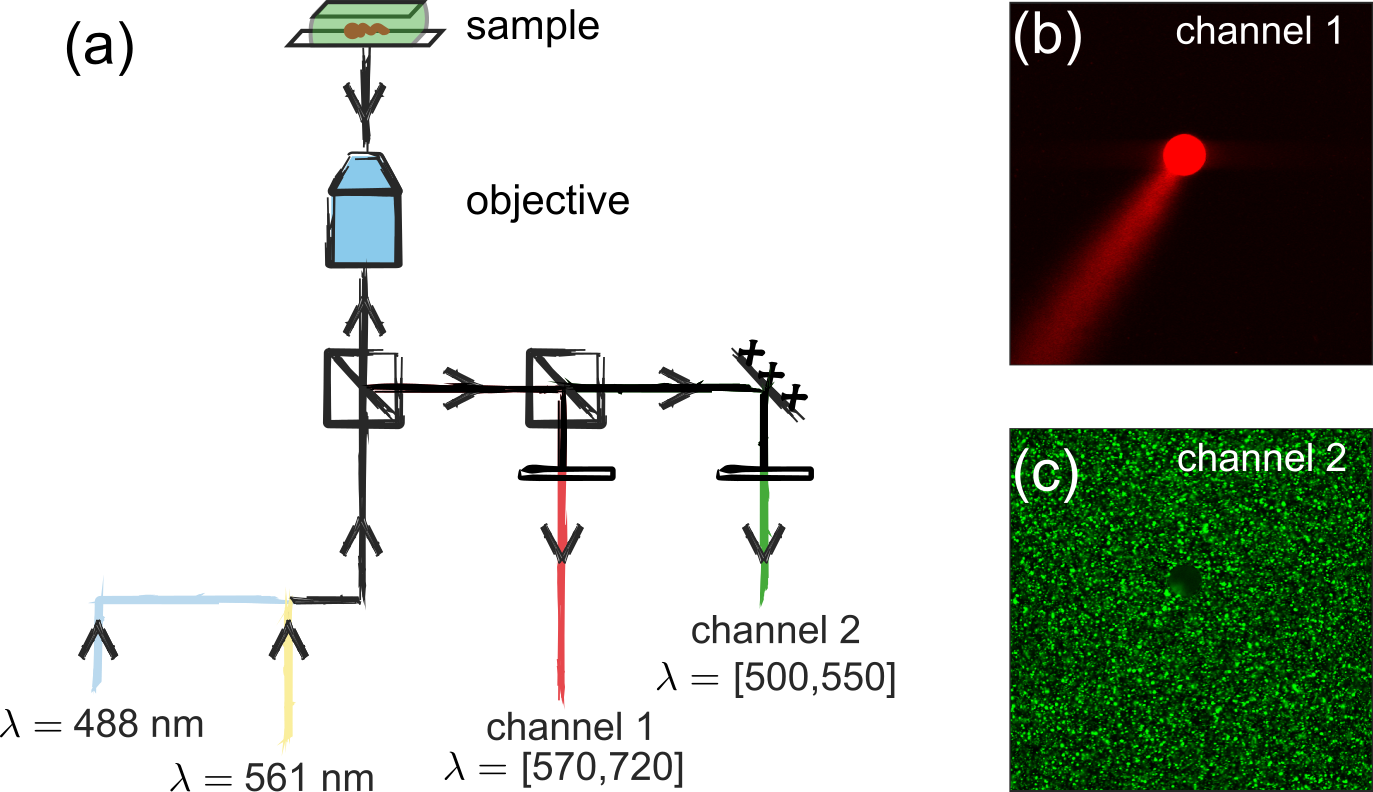}
\caption{Dual-channel fluorescent microscopy. \plb{a} Light path schematic with excitation laser lines. \plb{b, c} Example micrographs showing the separated emission from filled  micelles \plb{b} and fluorescent tracers \plb{c}.}\label{sifig:Technique}
\end{figure}

We used double-channel fluorescent microscopy for simultaneous imaging of the chemical and hydrodynamic fields. A schematic of the setup is shown in figure \figref{sifig:Technique}. Two laser units excite the test section. The Nile Red dye (Thermo Fisher Scientific), which visualises the oil phase, is excited with a 561\,nm laser and emits light at a maximum of $\sim$ 630\,nm. The green fluorescent particles (FluoSpheres$^{\mathrm{TM}}$, yellow-green fluorescent, 500\,nm in diameter), which visualise the fluid flow around the droplet, are excited with a 488\,nm laser and emit light at a maximum of $\sim 510$\,nm. The emitted light was separated using a beam splitter and appropriate filters for each emission maximum. We also used a spatial pinhole (confocal microscopy) to enhance image quality. Examples of snapshots recorded on each channel are shown in figure \figref[b,c]{sifig:Technique}.  

\subsection{Image processing and data analysis}\label{sisec:improc}
To observe the long time statistical behaviour of the active droplets, as in \figref{fig:statistics}, we observed their motion in a glass-bounded Hele-Shaw cell (quasi-two dimensional reservoir, $13\times8\,$mm and height
$h\approx60\, \mu$m) under a bright field microscope (Leica DM4000 B) at low magnification ($5\times$) compared to the double-channel fluorescence microscopy setup.
Videos were recorded at a frame rate of 10\,fps using a Canon (EOS 600d) digital camera ($1920\times1080$\,px). 
The droplet coordinates in each frame were extracted from video frames using the common Python libraries numpy, PIL and openCV (scripts available on request). 
Steps include background correction, binarisation, blob detection by contour analysis and minimum enclosing circle fits. 
Swimming trajectories were obtained using a frame-by-frame nearest-neighbour analysis.

To acquire the kymographs of the chemical field and tangential and radial velocities around the droplet interface, we observed the droplet behaviour by double-channel fluorescent microscopy as described in section \ref{sisec:DCFM}. 
We used a $512\times512$ pixels camera at a frame rate of 14 fps connected to a $20\times$ objective.
First we split the red (NileRed, filled micelles) and green (tracer particles) channels. 
Then, the red frames were used to extract the droplet coordinates via the blob detection algorithm described above. 
We developed a MATLAB script that centred the droplet and recorded the red light intensity value along the interface at a distance 15.6\,$\mu$m for $Pe\approx4$ and $36$ and 20.4\,$\mu$m for $Pe\approx293$. We note that it was not possible to record the intensity closer to the interface because the strong fluorescence from the large amounts of dye inside the droplet created a very bright region extending several micrometres beyond the actual interface.
We plotted the extracted profiles versus time to generate spatiotemporal kymographs. 

For a quantitative analysis of the flow field around the droplet we performed particle image velocimetry (PIV) on the tracer particles images (green channel) using the MATLAB-based  PIVlab interface~\cite{thielicke2014_pivlab}. 
The objective was focused on the mid-plane of the Hele-Shaw cell. We defined a moving mask for the area covered by the droplet.
We performed the analysis in $16\times16$ pixel interrogation windows with $75\%$ overlap. The spatial resolution is $1.2\;\mu \rm m/px$.
After obtaining the velocity vector field, we centered the droplet and read the velocity vectors at a certain distance from the droplet interface (3.6 $\mu m$ for $Pe\approx4$ and $36$ and 8.4 $\mu m$ for $Pe\approx293$). The tangential ($u_{\theta}$, in the droplet reference frame) and radial ($u_{r}$, only for $Pe\approx293$, in the lab reference frame) velocity components were then calculated and plotted in the kymographs. Due to the impermeability boundary condition, the radial component of the velocity directly at the interface is supposed to be zero; however, since we read the values at a certain distance from the interface there was an inward and outward radial contribution to the flow. We used this observation in particular at $Pe\approx293$ to show the quadrupolar symmetry of the flow field at the stopping moment.

In \figref[a]{fig:fillingcartoon} and supplementary video S1, we tracked the droplet and centred it in the image. To obtain the pathlines of the tracer particles in the video we used FlowTrace~\cite{gilpin2017_flowtrace} to convolve a superposition of 10 frames for each image. For \figref[a]{fig:fillingcartoon} we superimposed 30 frames. 
To visualise the motion of the tracer particles in \figref[b,IV]{fig:modestab} and the supplementary videos S6-S9, we processed the green channel of the input video (8 bit RGB) as follows: for each pixel coordinate, the intensity was replaced by its standard deviation within a 20 frame window around the current frame. Each frame was subsequently contrast maximised within a $[0,255]$ intensity range. The red and blue channels were not modified. This procedure was inspired by ImageJ's Z projection algorithm; the respective Python code is available on request. 

\section{Viscosity dependence of hydrodynamic modes}\label{sisec:hdmodel}
In this appendix we describe the mathematical framework for the coupled hydrodynamic and advection-diffusion problems pertaining to the active droplet system. Note that we have followed the solution methodology of refs. ~\cite{michelin2013_spontaneous,morozov2019_nonlinear, morozov2019_self-propulsion}, and have reworked each step of the analysis for the present system. The appendix shows the origins of all expressions and equations (including the scaling analyses necessary for simplifications) needed to understand the theoretical framework, and importantly, the origin of Fig. 4. We especially show each step of the linear stability analysis so that the derivation of the equations governing the instability growth rates for the hydrodynamic modes are clear. 
\subsection{Governing equations and boundary conditions for the active droplet system}\label{sisec:equations}
Considering \textit{an axisymmetric Stokes flow} (Reynolds no. for the swimming of the active droplet $Re\sim10^{-4}$), and \textit{the impermeability of the droplet interface}, the flow field around and inside the \textit{spherical} active droplet (capillary number $Ca<<1$) can be expressed in terms of the\textit{ non-dimensional} stream function $\psi$, in $(r, \; \theta)$ co-ordinate system, as ~\cite{leal2007_advanced,morozov2019_nonlinear,morozov2019_self-propulsion}: 
	
	\begin {equation} \label{eqn:psio}
	\begin {split}
		\psi^o&=a_1\left(\frac{1}{r}-r^2\right)(1-\eta^2)P_1^\prime(\eta)\\ &+\sum_{n=2}^{\infty}a_n\left(\frac{1-r^2}{r^n}\right)(1-\eta^2)P_n^\prime(\eta)
		\end {split}
	\end {equation} 
	\begin {equation} \label{eqn:psii}
	\psi^i=\sum_{n=1}^{\infty}b_n(r^{n+1}-r^{n+3})(1-\eta^2)P_n^\prime(\eta)
	\end {equation}
	
\noindent Here, and in the subsequent discussions, superscripts `\textit{o}' and `\textit{i}' refer to quantities outside and inside the active droplet respectively, $r$ is the radial coordinate non-dimensionalised by droplet radius $R_d$, $\eta =\cos \theta$, and $P_n(\eta)$ is the Legendre polynomial of degree $n$ with the prime denoting its derivative; $n$  here physically represents the $n^{th}$ hydrodynamic mode. 
The \textit{non-dimensional} radial and tangential flow velocity components around and inside the droplet are related to $\psi$ as $u_r=-\frac{1}{r^2}\frac{\partial \psi}{\partial \eta}$ and $u_\theta=-\frac{1}{r(1-\eta^2)^{1/2}}\frac{\partial \psi}{\partial r}$. The coefficients $a_n$ and $b_n$ in Eqs. \ref{eqn:psio} and \ref{eqn:psii} are constrained by the following boundary conditions	~\cite{leal2007_advanced,morozov2019_nonlinear}:	
	
\noindent	\textit{(i) tangential velocity $(u_{\theta})$ condition at the droplet interface $(r=1)$}:
	\begin {equation} \label{eqn:bcu}
	u_{\theta}^o-u_{\theta}^i=m \left(\frac{2\mu+3}{1+3m}\right)(1-\eta^2)^{1/2}\left(\frac{\partial c}{\partial\eta}\right)_{r=1}
	\end {equation} 
\noindent  \textit{(ii) tangential stress $(\tau_{r \theta})$ condition at the droplet interface $(r=1)$ (Marangoni effect)}: 
	\begin {equation} \label{eqn:bctau}
	\tau_{r\theta}^o-\tau_{r\theta}^i=-\frac{1}{\mu}\left(\frac{2\mu+3}{1+3m}\right)(1-\eta^2)^{1/2}\left(\frac{\partial c}{\partial\eta}\right)_{r=1}
	\end {equation}
	
\noindent The coefficients on the right hand side of Eqs. \ref{eqn:bcu} and \ref{eqn:bctau} essentially stem from the non-dimensionalization of the classical boundary conditions. 
Note that the flow velocity is non-dimensionalized using $V_t=\frac{q_s\left(\gamma_c R_d +3 \mu^i M \right)}{D\left(2\mu^o+3\mu^i\right)}$, which is a theoretical estimate for the terminal velocity of the active droplet considering the contributions of both the Marangoni and the diffusiophoretic effects~\cite{anderson1989_colloid,morozov2019_nonlinear}. 
Furthermore, $\mu = \mu^o/\mu^i$ is the ratio of the swimming medium viscosity $\mu^o$ to the droplet viscosity $\mu^i$, and the non-dimensional parameter $m$ represents the relative strengths of  diffusiophoretic to Marangoni effects ~\cite{morozov2019_nonlinear}. 
Essentially, $m$ can be considered as a ratio of the diffusiophoretic velocity scale to the viscocapilllary velocity scale representing the Marangoni effect.
Accordingly, $m=\frac{\mu^i M}{\gamma_c R_d} \approx  \frac{\zeta}{2R_d \mu}$, where $M \approx  \frac{k_BT}{2 \mu^o} \zeta^2$ is the diffusiophoretic mobility ~\cite{anderson1989_colloid, izri2014_self-propulsion}, $\gamma_c  \approx k_B T \zeta$ is the leading order change in the interfacial surface tension $\gamma$ with surfactant concentration $c$ (alternatively, $\gamma_c=\frac{d\gamma}{dc}$ can be considered to be a measure of the change in $\gamma$ with $c$ assuming a linear variation) ~\cite{izri2014_self-propulsion, morozov2019_nonlinear}, and $\zeta \sim 10$ nm is the characteristic length scale over which the surfactants interact with the droplet in the interfacial region.    
For the active droplet system, $O(m) \sim 10^{-3}-10^{-2}$ for the entire range of experiments; hence, for the present physical problem the diffusiophoretic effect is much weaker as compared to the Marangoni effect. However, the former is considered in the analysis here for the sake of generality.
In the definition of $V_t$, $q_s$ is an isotropic and constant interfacial surfactant consumption rate per unit area necessary for the droplet activity, and $D=\frac{k_BT}{6 \pi r_s \mu^o}$ is the diffusion coefficient for the surfactant monomer (length scale for surfactant monomer $r_s \sim 10^{-10}$ m). $q_s$ can be approximately estimated by assuming that the total number of surfactant monomers necessary per unit time to account for the volumetric reduction rate of the droplet due to the formation of the filled micelles is equal to the total interfacial surfactant consumption rate. Hence, $|dV_d/dt|N_s/v_{fm} \approx q_s 4 \pi R_d^2$, which implies that $q_s \approx (3 N_s |dR_d/dt|)/(4 \pi r_{fm}^3)$. Here, $O(N_s) \sim 25$ is the number of surfactant monomers per filled micelle, $v_{fm}=4/3 \pi r_{fm}^3$ is the filled micelle volume with a micellar radius of $O(r_{fm}) \sim 2.5$ nm, and $|dR_d/dt|$ is the droplet solubilization rate as given in Fig. \ref{sifig:Solubilisation}. \\

Eqs. \ref{eqn:bcu} and \ref{eqn:bctau} delineate the dependence of the swimming hydrodynamics on the distribution of the non-dimensional surfactant concentration $c$ in the  vicinity of the droplet. 
Naturally, $c$ is governed by an advection-diffusion relation~\cite{leal2007_advanced,morozov2019_nonlinear,morozov2019_self-propulsion}:

	\begin {equation} \label{eqn:advdiff}
	\begin{split}
		&Pe\left[u_r^o\frac{\partial c}{\partial r}- \frac{u_{\theta}^o}{r}(1-\eta^2)^{1/2} \frac{\partial c}{\partial \eta}\right]\\
		&=\frac{1}{r^2} \frac{\partial}{\partial r}\left(r^2\frac{\partial c}{\partial r}\right )+\frac{1}{r^2} \frac{\partial}{\partial \eta}\left((1-\eta^2)\frac{\partial c}{\partial \eta}\right)
	\end{split}
	\end {equation}  
	
\noindent The distribution of $c$ is subject to the following boundary conditions:
	
\noindent \textit{(i) isotropic and constant surfactant consumption at the droplet interface (r=1)}
   \begin {equation} \label{eqn:bcsurfcons}
	\left( \frac{\partial {c}}{\partial r} \right)_{r=1}=1
	\end {equation}
	
\noindent \textit{(ii) the bulk condition}
   \begin {equation} \label{eqn:bcbulk}
	 c(r \rightarrow \infty)\rightarrow c_{\infty}
	\end {equation}
	 
\noindent Note that Eq. \ref{eqn:bcsurfcons} addresses the depletion of the interfacial surfactant monomers due to the creation of the filled micelles by considering the isotropic and constant interfacial surfactant adsorption rate per unit area of $q_s$, corresponding to a flux with unit of number per area per time (in dimensional form: $D \nabla c^* \cdot \hat{n}= q_s$; this gives a scale for the surfactant concentration as $\sim \frac{q_s R_d}{D}$) ~\cite{morozov2019_nonlinear, morozov2019_self-propulsion}. 
$Pe$ in Eq. \ref{eqn:advdiff} is the system P\'eclet number-- the details of which are discussed in the following sub-section. 
The above system of equations (Eqs. \ref{eqn:psio}--\ref{eqn:bcbulk}) can be solved for $\psi$ (therefore $u_r$, $u_\theta$), and $c$ using the singular perturbation technique for certain limiting cases  ~\cite{morozov2019_nonlinear, morozov2019_self-propulsion}. 
The solvability condition clearly shows that the actuations of different hydrodynamic modes  depend on certain threshold values of $Pe$ (Fig. 4a in the main text) ~\cite{ morozov2019_nonlinear}. 
Furthermore, the asymptotic analysis also provides a physical understanding of the hydrodynamic and surfactant concentration fields corresponding to the different modes, specifically $n=1$ and $n=2$ (Fig. 4b in the main text). \\
	
\subsection{The system P\'eclet number}\label{sisec:peclet}
The important thing to understand now is the dependence of $Pe$ on $\mu$. 
Classically, $Pe$ can be written as $Pe=\frac{V_t R_d}{D}$, where $V_t=\frac{q_s\left(\gamma_c R_d +3 \mu^i M \right)}{D\left(2\mu^o+3\mu^i\right)}$ is the theoretical estimate for the terminal velocity of the active droplet considering the contributions of both the Marangoni and diffusiophoretic effects, as mentioned in the preceding sub-section ~\cite{anderson1989_colloid,morozov2019_nonlinear}. Utilizing the aforementioned definition of $V_t$, and following some simple algebraic manipulations, $Pe$ can be expressed in terms of system constants and the parameter $\mu$ as:

\begin {equation} \label{eqn:pes}
\begin{split}
	&Pe=\frac{V_t R_d}{D}=\frac{q_s\left(\gamma_c R_d +3 \mu^i M \right)}{D\left(2\mu^o+3\mu^i\right)}\frac{R_d}{D}\\
	&\Rightarrow Pe=\frac{q_s M}{Dm} \frac{\left(1+3m\right)}{\left(2\mu+3\right)}\frac{R_d}{D}\\
	&\Rightarrow Pe \approx \frac{18 \pi^2}{k_BT}  q_s r_s^2 \zeta R_d^2 \mu^i \left[\mu \left(\frac{2\mu+3 \zeta/R_d}{2 \mu+3}\right)\right]\\
 \end{split}
\end {equation}
 
 \noindent In the last step of Eq. \ref{eqn:pes}, the approximate expressions for $M$ and $m$ (see sub-section~\ref{sisec:equations}), and the definition of $D$ (see sub-section~\ref{sisec:equations}) are utilized to derive the final expression for $Pe$. 
 Eq. \ref{eqn:pes} expresses $Pe$ as a monotonically increasing function of the viscosity ratio $\mu$ (markers in Fig. 4a in the main text). 
 Note that $q_s$ is approximately estimated by relating the dissolution rate of the active droplet to the isotropic and constant surfactant consumption at the droplet interface ~\cite{izri2014_self-propulsion}; the dissolution rate of the active droplet is dependent on the glycerol concentration (\figref{sifig:Solubilisation}) which effectively makes $q_s$ dependent on $\mu_o$ . We further note that the second term in the numerator within parenthesis $O\left(\frac {\zeta} {R_d} \right) \sim 10^{-4}$; this further substantiates the fact that the diffusiophoretic effect is much weaker compared to the Marangoni effect for the present system.\\    
 
\subsection{Linear stability analysis about a motionless (isotropic) base state}\label{sisec:linstab}
For the linear stability analysis (also see~\cite{michelin2013_spontaneous, morozov2019_self-propulsion}), the time-dependent form of the advection-diffusion equation (Eq. \ref{eqn:advdiff}) is used: 

\begin {equation} \label{eqn:advdiff_unsteady}
	\begin{split}
		&Pe\left[\frac{\partial c}{\partial t}-\frac{1}{r^2} \frac{\partial \psi^o }{\partial \eta} \frac{\partial c}{\partial r}+ \frac{1}{r^2} \frac{\partial \psi^o }{\partial r} \frac{\partial c}{\partial \eta}\right]\\
		&=\frac{1}{r^2} \frac{\partial}{\partial r}\left(r^2\frac{\partial c}{\partial r}\right )+\frac{1}{r^2} \frac{\partial}{\partial \eta}\left((1-\eta^2)\frac{\partial c}{\partial \eta}\right)
	\end{split}
\end {equation}

\noindent Next, the desired quantities are expressed in terms of the unsteady (instability) modes-- $\psi=e^{\lambda t} \sum_{n} \tilde{\psi}_n(r) P_n(\eta)$ and $c=-\frac{1}{r}+e^{\lambda t} \sum_{n} \tilde{c}_n(r) P_n(\eta)$, where $\lambda(>0)$ is the non-dimensional growth rate for the instability modes. 
Using the aforementioned expressions for $\psi$ and $c$, and linearizing Eq. \ref{eqn:advdiff_unsteady}, the governing equations for the first two modes can be obtained as:
  
\begin {equation} \label{eqn:c_mode1}
\frac{d}{dr}\left(r^2 \frac{d \tilde{c}_1}{dr}\right)- \left( 2+\lambda_s^2 r^2 \right) \tilde{c}_1 = 2 Pe \; a_1 \frac{1-r^3}{r^3}	
\end {equation} 
 
\begin {equation} \label{eqn:c_mode2}
\frac{d}{dr}\left(r^2 \frac{d \tilde{c}_2}{dr}\right)- \left( 6+\lambda_s^2 r^2 \right) \tilde{c}_2 = 6 Pe \; a_2 \frac{1-r^2}{r^4}	
\end {equation}

\noindent where $\lambda_s=\sqrt{\lambda \; Pe}$, and $a_1$ and $a_2$ are the coefficients of the first and second modes respectively of the outer stream function (as in Eq. \ref{eqn:psio}). Eqs. \ref{eqn:c_mode1} and \ref{eqn:c_mode2} are solved to evaluate $\tilde{c}_1$   and $\tilde{c}_2$, respectively:

\begin {equation} \label{eqn:c_1}
\begin {split}
\tilde{c}_1&= Pe \; a_1 \left( \frac{2}{x^2} + \frac {\lambda_s^3}{2 x^3} \right) + \alpha_1 \left(\frac{1+x}{2 x^2}\right) e^{-x}\\
                    &- Pe \; a_1  \frac {\lambda_s^3}{4 x^2} \left[ \left(\frac{1+x}{2}\right) \left (Chi(x)+Shi(x) \right) e^{-x}\right. \\
                    &- \left. \left(\frac{1-x}{2}\right) \left (Chi(x)-Shi(x) \right) e^{x}  \right]
\end {split}
\end {equation} 

\begin {equation} \label{eqn:c_2}
\begin {split}
\tilde{c}_2&= Pe \; \frac{a_2}{8} \left( \frac{8 \lambda_s^4}{x^4} + \frac{\lambda_s^4}{ x^2}  - \frac{6 \lambda_s^2}{x^2}       \right) + \alpha_2 \left(\frac{x^2+3x+3}{2 x^3}\right) e^{-x}\\
                &+ Pe \; a_2  \frac{\lambda_s^2 (6-\lambda_s^2)}{16 x^3} \left[\left(x^2+3x+3\right) \left(Chi(x)+Shi(x) \right) e^{-x}\right.\\
                &- \left.\left(x^2-3x+3\right) \left(Chi(x)-Shi(x) \right) e^{x}\right]
\end {split}
\end {equation} 

\noindent Here, $x=r\lambda_s$ is a rescaled spatial variable, $Chi(x)$ and $Shi(x)$ are the hyperbolic cosine integral and hyperbolic sine integral functions, and $\alpha_1$ and $\alpha_2$ are the constants of integration. 
Note that Eqs. \ref{eqn:c_1} and \ref{eqn:c_2} are evaluated in a manner which satisfies the bulk condition for the surfactant distribution (Eq. \ref{eqn:bcbulk}) i.e. as $r \rightarrow \infty$, $c-c_{\infty}\rightarrow 0$. 
Furthermore, considering the expression for $c$, the interfacial surfactant consumption condition (Eq. \ref{eqn:bcsurfcons}) reduces to the form:

\begin {equation} \label{sufcon_mod}
\begin {split}
	&\left( \frac{\partial {c}}{\partial r} \right)_{r=1}=1 \Rightarrow \left(\frac{d\tilde{c}_1}{dx}\right)_{x=\lambda_s}=0; \; \left(\frac{d\tilde{c}_2}{dx}\right)_{x=\lambda_s}=0
\end {split}
\end {equation} 

\noindent Using Eqs. \ref{eqn:c_1} and \ref{sufcon_mod}, $\alpha_1$ can be evaluated as: 

\begin {equation} \label{eqn:alpha_1}
\begin {split}
\alpha_1&=-Pe \; a_1 \frac{e^{\lambda_s}}{4\left(\lambda_s^2+2\lambda_s+2\right)} \left[2\left(\lambda_s^4+6\lambda_s^2+16\right)\right.\\
&+ \lambda_s^3 \left(\lambda_s^2-2\lambda_s+2\right) \left(Chi(\lambda_s)-Shi(\lambda_s) \right) e^{\lambda_s} \\
&-  \left.\lambda_s^3 \left(\lambda_s^2+2\lambda_s+2\right) \left(Chi(\lambda_s)+Shi(\lambda_s) \right) e^{-\lambda_s}\right]
\end {split}
\end {equation} 

\noindent Similarly, using Eqs. \ref{eqn:c_2} and \ref{sufcon_mod}, $\alpha_2$ can be evaluated as: 

\begin {equation} \label{eqn:alpha_2}
\begin {split}
\alpha_2&=Pe \; a_2 \frac{\lambda_s^2 e^{\lambda_s}}{8\left(\lambda_s^3+4\lambda_s^2+9\lambda_s+9\right)} \left[-2 \lambda_s \left(5\lambda_s^2+2\right)\right.\\
&+ \left(\lambda_s^2-6\right) \left(\lambda_s^3-4\lambda_s^2+9\lambda_s-9\right) \left(Chi(\lambda_s)-Shi(\lambda_s) \right) e^{\lambda_s}\\
&+ \left. \left(\lambda_s^2-6\right) \left(\lambda_s^3+4\lambda_s^2+9\lambda_s+9\right) \left(Chi(\lambda_s)+Shi(\lambda_s) \right) e^{-\lambda_s}\right]
\end {split}
\end {equation}

\noindent Eqs. \ref{eqn:c_1} and \ref{eqn:c_2}, along with Eqs. \ref{eqn:alpha_1} and \ref{eqn:alpha_2}, give closed form expressions for $\tilde{c}_1$ and $\tilde{c}_2$.\\

Considering the hydrodynamic boundary conditions (Eqs. \ref{eqn:bcu} and \ref{eqn:bctau}), and using the orthogonality condition for Legendre polynomials, a set of two simple algebraic equations  for the co-efficients $a_n$, and $b_n$ for each of the first two modes can be written as:

\noindent \textit{(i) first mode $(n=1)$}
\begin {equation} \label{eqn:solva_1a}
3a_1-2b_1=m \left(\frac{2\mu+3}{1+3m}\right) \tilde{c}_1
\end {equation}
\begin {equation} \label{eqn:solva_1b}
\mu a_1+b_1=\frac{1}{6}\left(\frac{2\mu+3}{1+3m}\right) \tilde{c}_1
\end {equation}

 \noindent \textit{(ii) second mode $(n=2)$}
 \begin {equation} \label{eqn:solva_2a}
 a_2-b_2=\frac{m}{2} \left(\frac{2\mu+3}{1+3m}\right) \tilde{c}_2
 \end {equation}
 \begin {equation} \label{eqn:solva_2b}
 \mu a_2+b_2=\frac{1}{10}\left(\frac{2\mu+3}{1+3m}\right) \tilde{c}_2
 \end {equation}
 
\noindent Note that $\tilde{c}_n$ in the above equations is explicitly dependent on $a_n$ (see Eqs. \ref{eqn:c_1}, \ref{eqn:alpha_1} and \ref{eqn:c_2}, \ref{eqn:alpha_2}). 
Considering the closed form expression for $\tilde{c_1}$ (Eq. \ref{eqn:c_1} and \ref{eqn:alpha_1}), the solvability condition for Eq. \ref{eqn:solva_1a} and \ref{eqn:solva_1b} gives:

\begin {equation} \label{eqn:ins_s1}
\frac{-\left(Chi(\lambda_s)-Shi(\lambda_s)\right) e^{\lambda_s} \lambda_s^4 - \lambda_s^3 + \lambda_s^2 -2\lambda_s +6}{12 \left(\lambda_s^2+2 \lambda_s +2\right)} = \frac {1}{Pe}
\end {equation}

\noindent Similarly, considering the closed form expression for $\tilde{c_2}$ (Eq. \ref{eqn:c_2} and \ref{eqn:alpha_2}), the solvability condition for Eq. \ref{eqn:solva_2a} and \ref{eqn:solva_2b} gives: 
\begin{equation} \label{eqn:ins_s2}
\begin{split}
&\frac{-(6-\lambda_s^2) \left(Chi(\lambda_s)-Shi(\lambda_s)\right) e^{\lambda_s} \lambda_s^4}{8  \left(\lambda_s^3+4\lambda_s^2+9\lambda_s+9\right)} \\
+& \frac{\left(\lambda_s^2+2 \lambda_s +2\right) \left(\lambda_s^3-3 \lambda_s^2 +6\right)}{8  \left(\lambda_s^3+4\lambda_s^2+9\lambda_s+9\right)} \\
=& \frac {10}{Pe} \frac{(1+\mu)(1+3m)}{(2\mu+3)(1+5m)}
\end{split}
\end{equation}
\noindent Eqs. \ref{eqn:ins_s1} and \ref{eqn:ins_s2} are solved numerically to evaluate the variations of the non-dimensional growth rates $\left(\lambda=\frac{\lambda_s^2}{Pe} \right)$ with $Pe$ for the first and second instability modes respectively (Fig. 4c in the main text). Note that Eq. \ref{eqn:ins_s1} is identical to that derived for the spontaneous motion of an autophoretic isotropic particle ~\cite{michelin2013_spontaneous}. Furthermore, it is important to note here that the inverse of the time scale used for non-dimensionalizing the growth rate is $\frac{V_t}{R_d}$, which is consistent with the entire analysis.  

\section{Supplementary figures}\label{appsec:figures}
\begin{figure}
\centering\vspace{5mm}\includegraphics[width=.7\columnwidth]{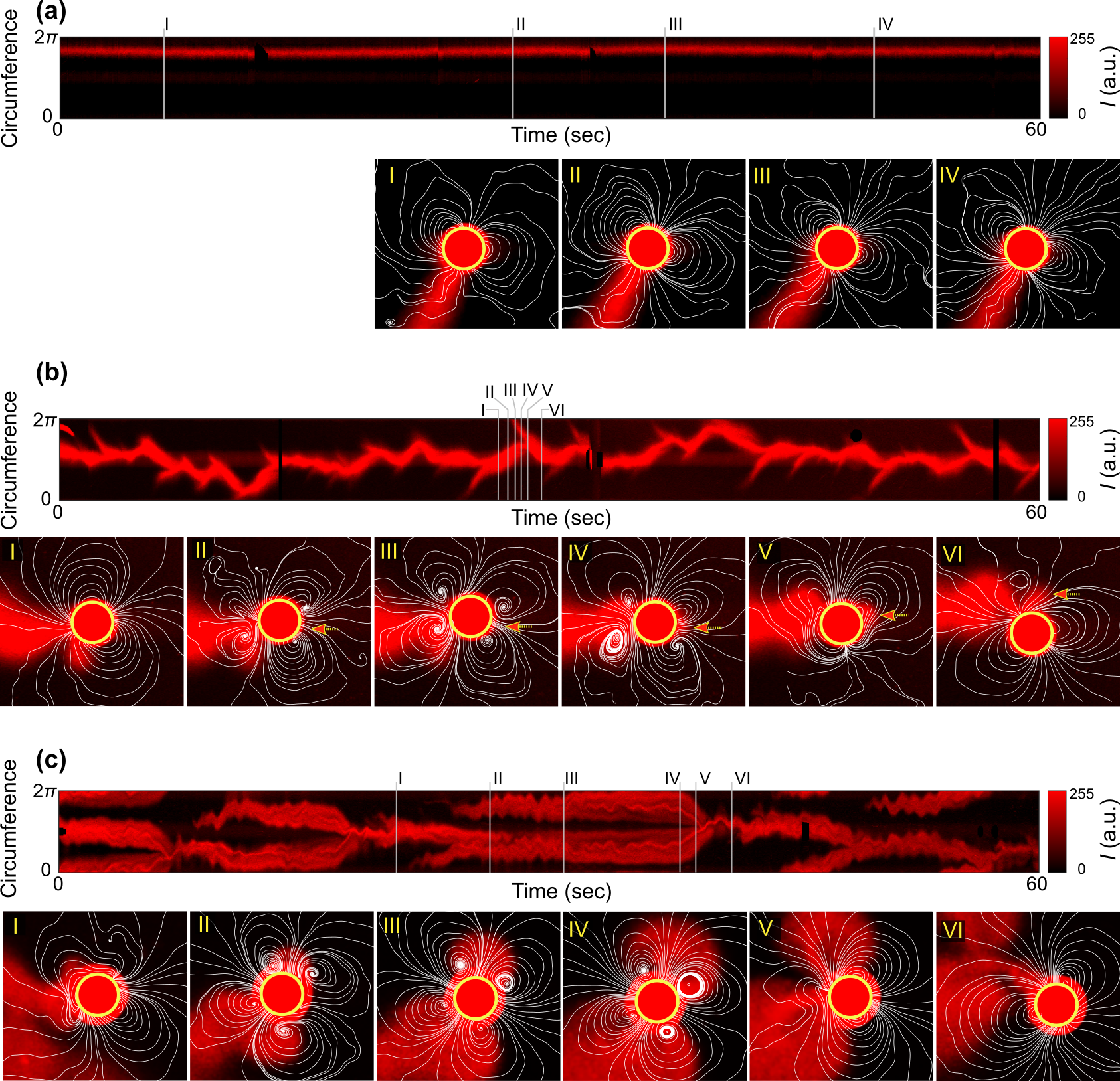}
\caption{Additional data for \figref{fig:kymographs}: Kymographs of the chemical field with selected instantaneous frames. Rows \plb{a}, \plb{b} and \plb{c} respectively correspond to $Pe=4$, $Pe=36$ and $Pe=293$. In \plb{b}, the red arrow shows the location of the growing filled micelle blob. }\label{sifig:ComplementaryFig4}
\end{figure}

Extending \figref{fig:kymographs}, \figref{sifig:ComplementaryFig4} provides additional flow field snapshots to illustrate transient flow modes,  with the chemical field kymographs plotted for a longer period of 60 seconds.  Supporting Videos S6-8 respectively correspond to the kymographs in \figref{sifig:ComplementaryFig4}(a-c).   

\begin{figure}
\centering\includegraphics[width=.7\columnwidth]{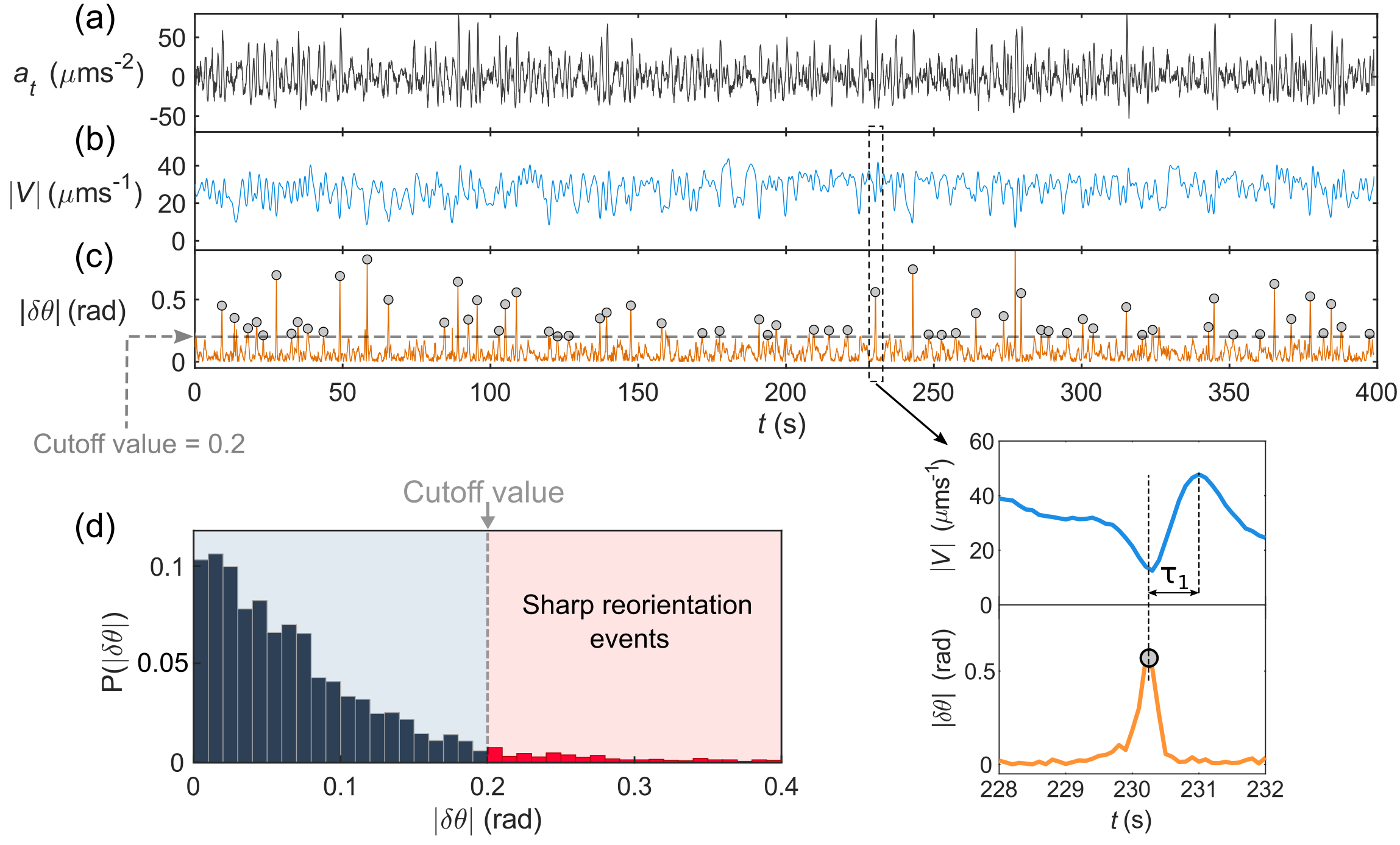}
\caption{Conditional averaging over reorientation events. \plb{a}, \plb{b} and \plb{c} are tangential acceleration, speed and the reorientation angle magnitude, respectively. In \plb{c}, the identified sharp reorientation events are shown by grey ($\circ$) symbols. The zoomed-in view is one example event that shows the general trend, a delay, $\tau_1$, between $|\delta\theta|$ and $|V|$. \plb{d} The distribution of $|\delta\theta|$ and the cutoff value. The sharp turning events are coloured in red.}\label{sifig:Cutoff}
\end{figure}

In \figref{sifig:Cutoff}, we have plotted the long-time tangential acceleration, speed and the reorientation angle for $Pe=36$. This data set was used to identify the abrupt reorientation events.  
We identified these events based on a cutoff criterion for the reorientation between video frames $|\delta\theta| = 0.2$ rad (\figref{sifig:Cutoff}, c \& d), aligned and overlaid the profiles of all events with the turning point ($|\delta\theta_{max}|$) set as $t = 0$, and calculated the time-dependent average ($\langle{}\rangle$ represents ensemble averaging over all events).

\begin{figure*}
\centering\includegraphics[width=.8\textwidth]{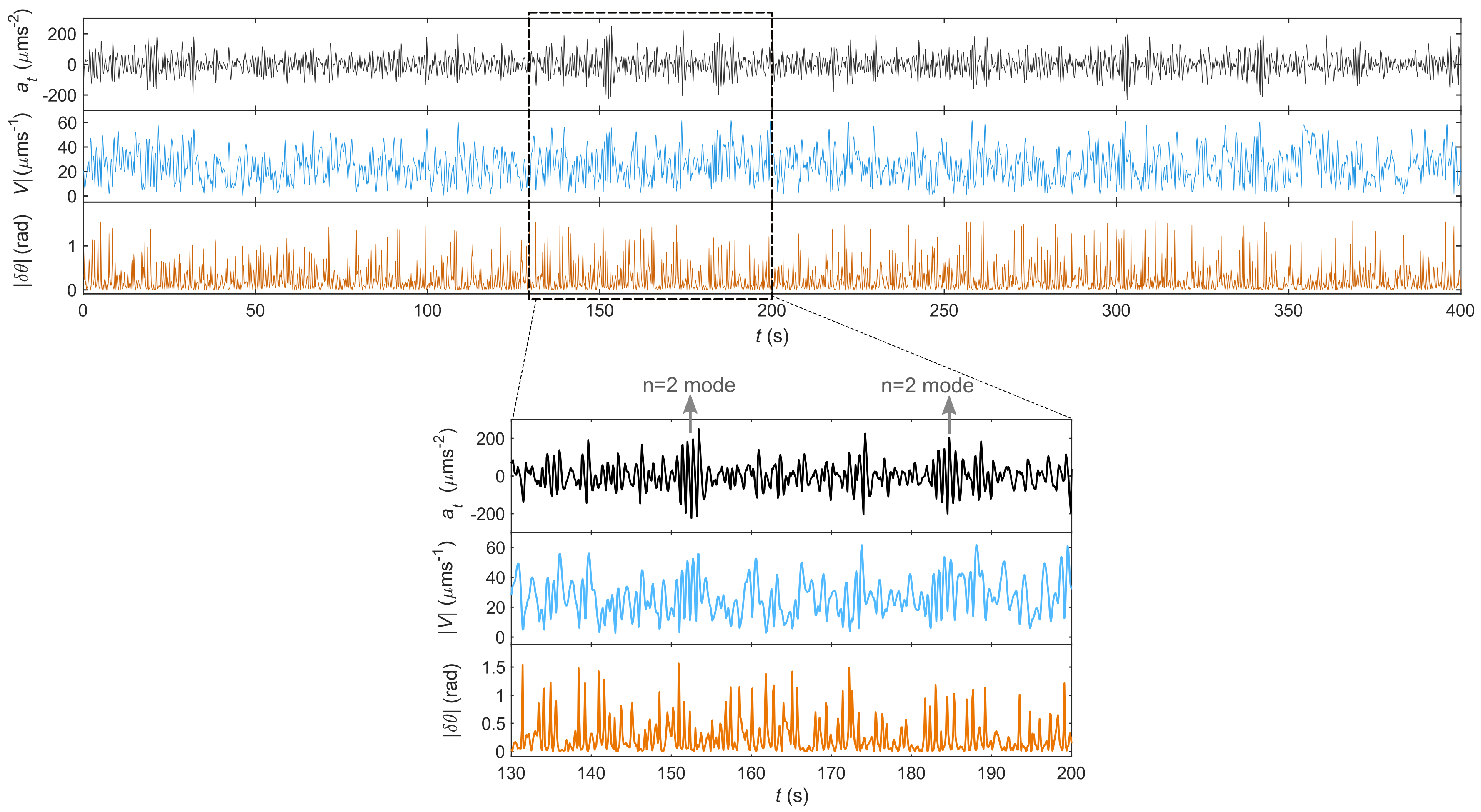}
\caption{Signatures of bimodal space exploration in the long-time tangential acceleration signal. The corresponding $Pe$ is 293.}\label{sifig:BimodalSignal}
\end{figure*}

In \figref{sifig:BimodalSignal}, we plotted the the long time acceleration signal for $Pe=293$ to demonstrate signatures of bimodal swimming.  Such events can be identified by intermittent strong fluctuations in the acceleration profile. The zoomed-in view further demonstrates the difference between stopping ($n=2$) and swimming modes ($n=1$). Constant transitions beween these modes result in the anomalous diffusive behaviour shown in \figref{fig:SAW} in the main text.

 \section{Supplementary movies}
\begin{figure*}
 \begin{minipage}{\textwidth}
\includegraphics[width=.5\columnwidth]{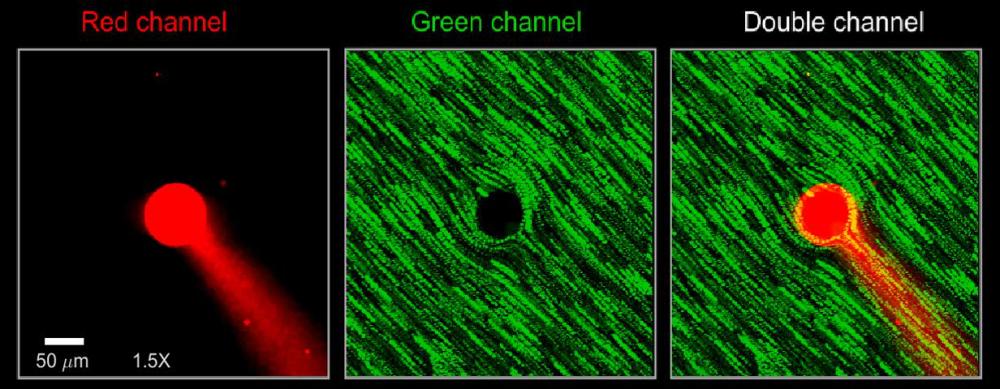}
\caption{{\bf Supplementary Video S1}. Simultaneous visualisation of hydrodynamic and chemical field. The videos from splitting colour channels (red and green) and the composite image obtained through double-channel fluorescent microscopy. The droplet is tracked and centred. The particle pathlines are obtained by superposition of 10 frames for each image. The video is played $1.5\times$ faster than the real time.}\label{movie:S1}
\end{minipage}\hspace{\stretch{1}}%
\end{figure*}

\begin{figure*}
\begin{minipage}{\textwidth}
\includegraphics[width=.24\columnwidth]{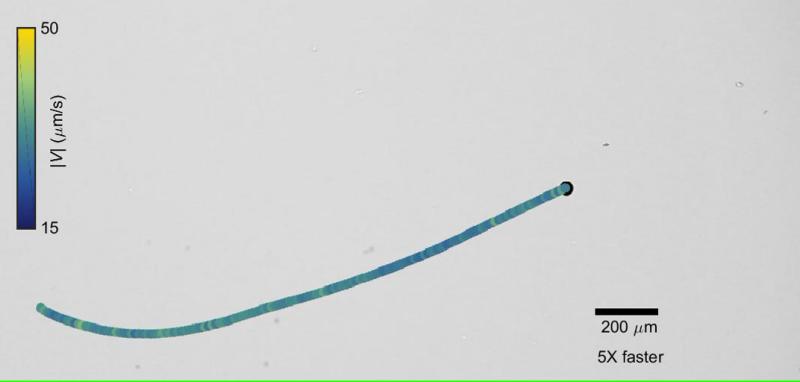}\hspace{\stretch{1}}
\includegraphics[width=.24\columnwidth]{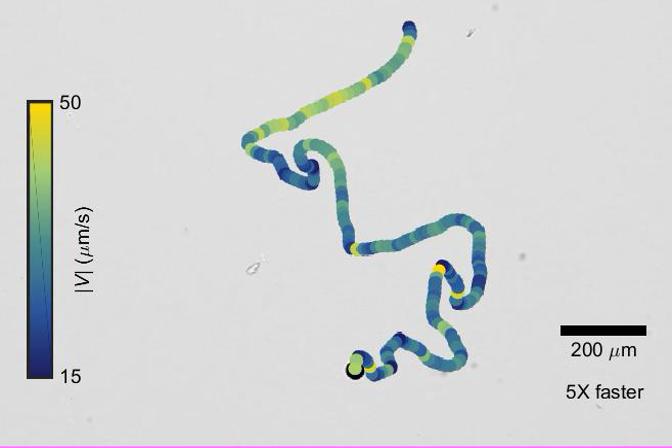}
\includegraphics[width=.24\columnwidth]{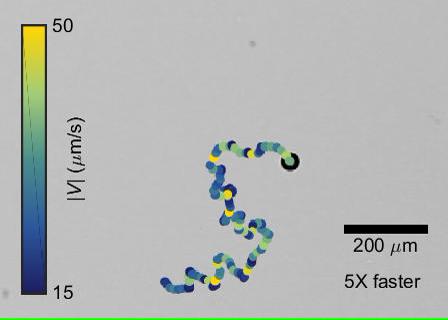}\hspace{\stretch{1}}
\includegraphics[width=.24\columnwidth]{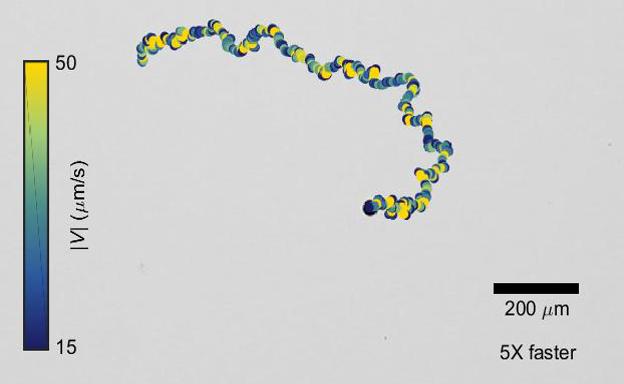}\\
\caption{{\bf Supplementary Video S2-S5}. The trajectories colour-coded with speed value. From left to right S2, S3, S4 and S5 correspond to $Pe=4$ ($\mu^o=1.2$ mPa.s), $Pe=19$ ($\mu^o=2.6$ mPa.s), $Pe=101$ ($\mu^o=5.5$ mPa.s) and $Pe=1112$ ($\mu^o=17.4$ mPa.s). The increase in the swimming medium viscosity destabilises the propulsion dynamics. The videos are played $5\times$ faster than the real time.}\label{movie:S3}
\end{minipage}\hspace{\stretch{1}}%
\end{figure*}

\begin{figure*}
\begin{minipage}{\textwidth}
\includegraphics[width=.32\columnwidth]{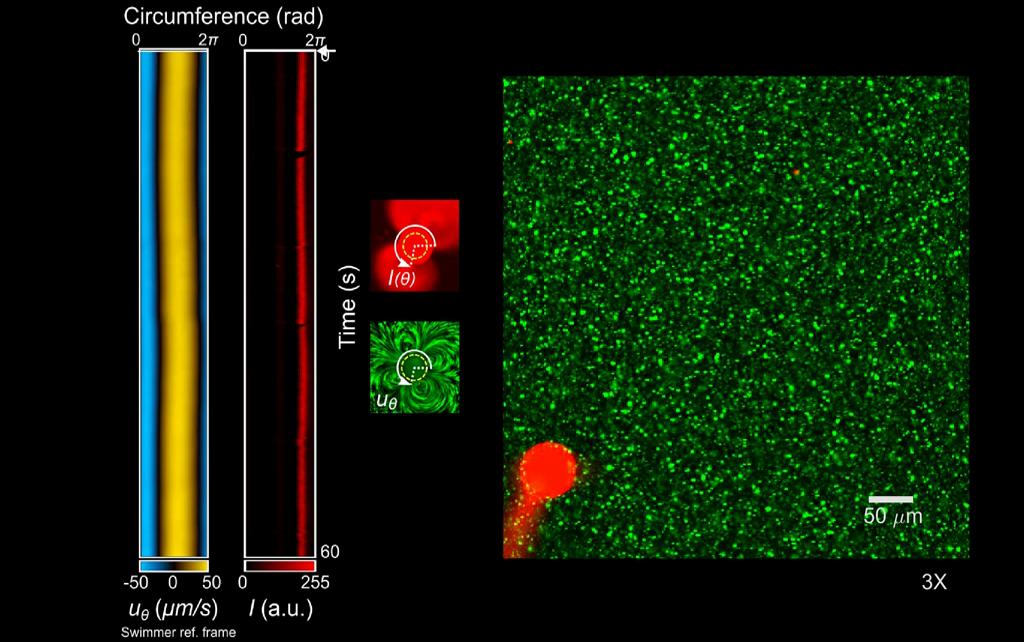}\hspace{\stretch{1}}
\includegraphics[width=.32\columnwidth]{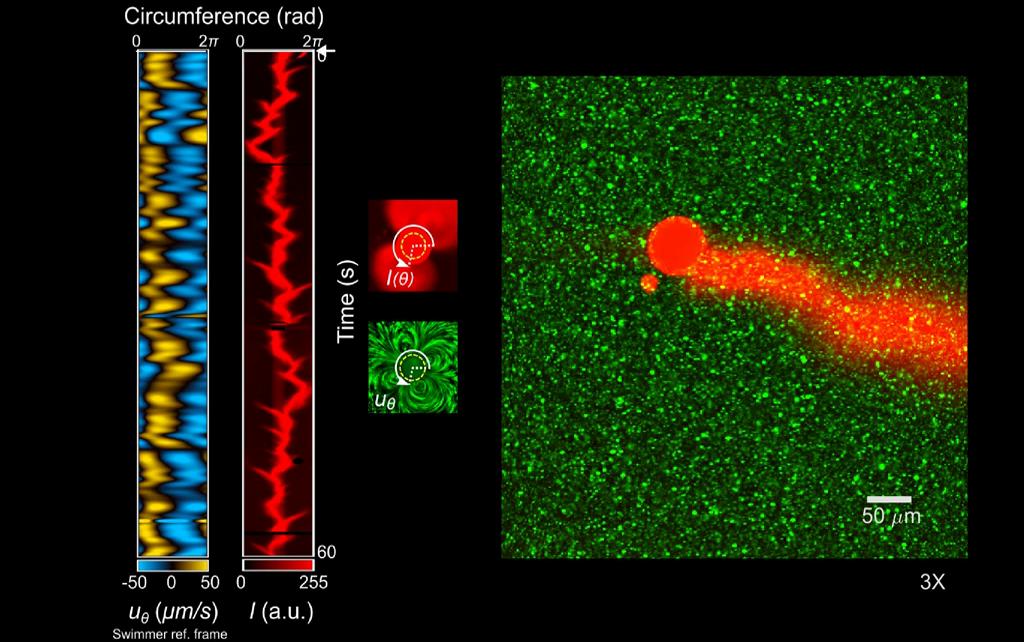}\hspace{\stretch{1}}
\includegraphics[width=.32\columnwidth]{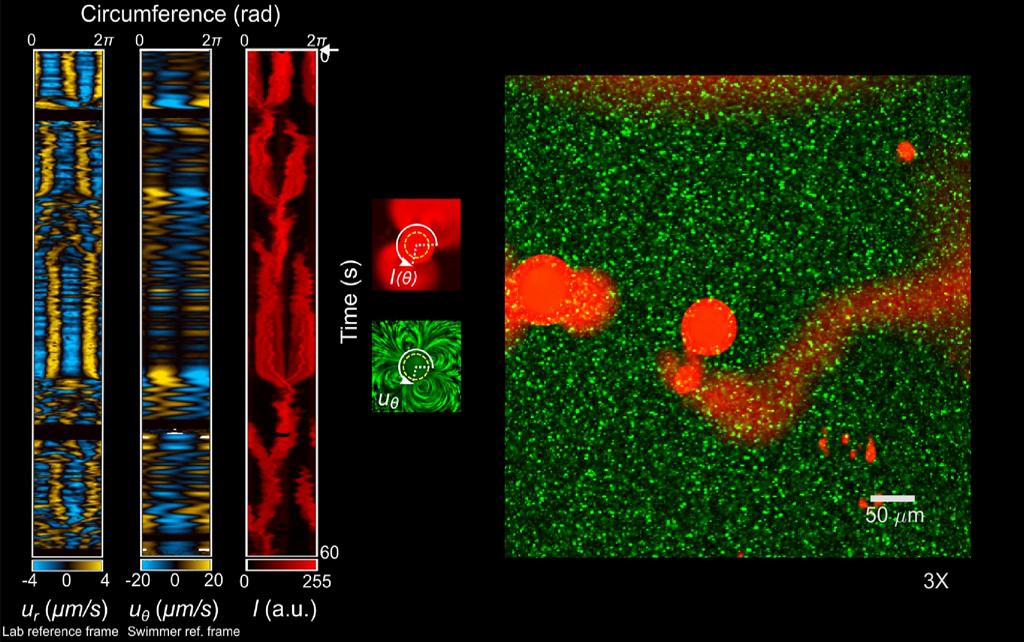}
\caption{{\bf Supplementary Video S6-S8}. The kymographs of chemical fields ($I(\theta)$), tangential  velocity component ($u_{\theta}$) and radial velocity component ($u_r$, only for $Pe=293$). From left to right S6, S7 and S8 correspond to $Pe=4$, $Pe=36$ and $Pe=293$. The particle pathlines are obtained by superposition of 20 frames for each image. Interactions with solubilisation history results in the emergence of unsteady spatiotemporal dynamics. The videos are played $3\times$ faster than the real time.}\label{movie:S6}
\end{minipage}\hspace{\stretch{1}}%
\end{figure*}

\begin{figure*}
\begin{minipage}{\textwidth}
\centering\includegraphics[width=.49\columnwidth]{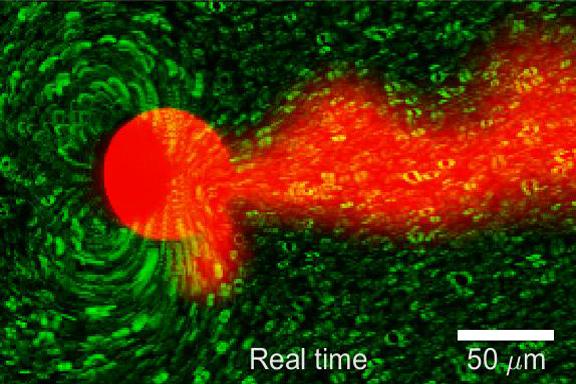}
\caption{{\bf Supplementary Video S9}. Growth and accumulation of filled micelle blob in the leading front of the droplet resulting in sharp reorientation. The corresponding $Pe$ is 36. First, the red channel is shown and later the green channel with the tracer particles pathlines are superimposed. The short-lived appearance of the quadrupolar flow field is followed by transitioning to the dipolar flow field.}\label{movie:S9}
\end{minipage}\hspace{\stretch{1}}%
\end{figure*}

\begin{figure*}
\begin{minipage}{\textwidth}
\centering\includegraphics[width=.6\columnwidth]{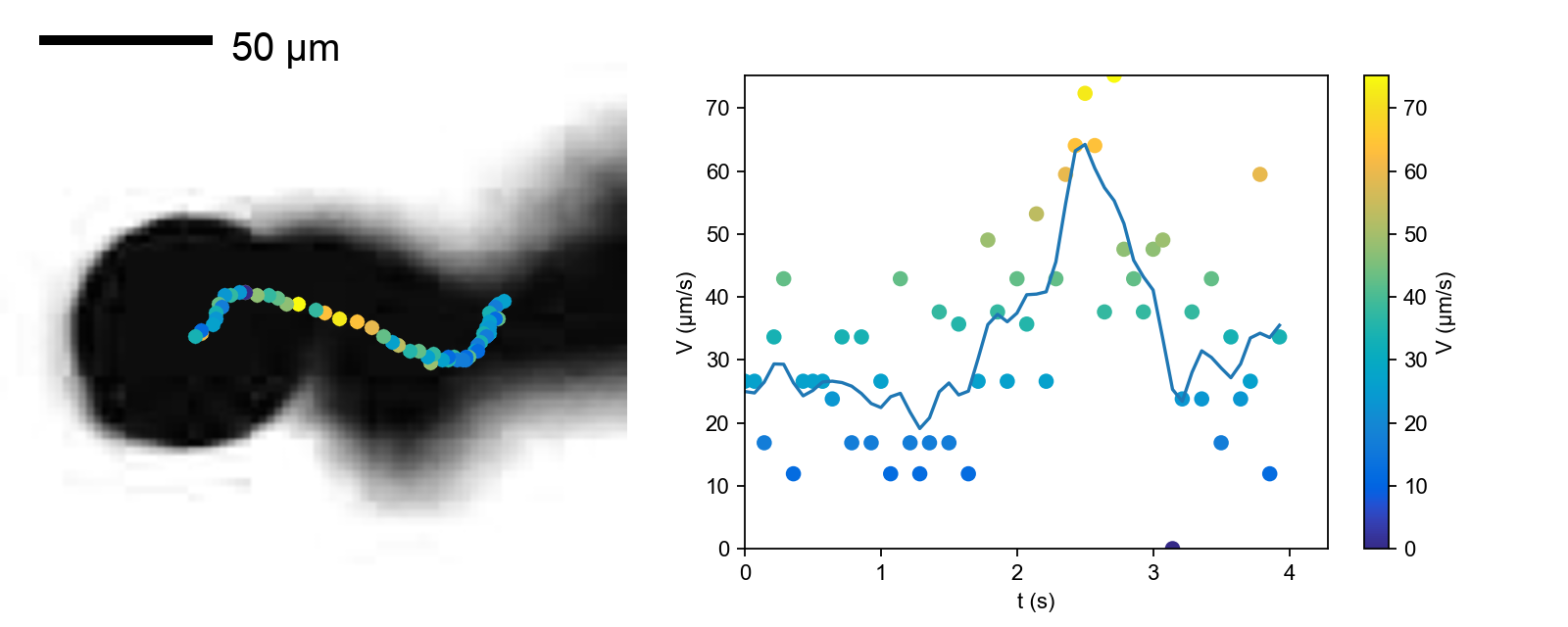}
\caption{{\bf Supplementary Video S10}. Temporal variation of propulsion speed during a sharp reorientation event, due to interaction with the secondary filled micelles aggregate. The trajectory of the droplet is colour-coded with instantaneous speed. The corresponding $Pe$ is 36.}\label{movie:S10}
\end{minipage}\hspace{\stretch{1}}%
\end{figure*}

\figref{movie:S1}-\figref{movie:S10} provide thumbnail previews and explanatory captions for the supplementary movies (deposited under \url{http://asm.ds.mpg.de/index.php/media/#stopandgo2020}).

 \cleardoublepage
 \onecolumngrid
 
\end{document}